\documentclass[journal,onecolumn,web]{ieeecolor}

\usepackage{generic}
\usepackage{cite}
\usepackage{amsmath,amssymb,amsfonts}
\usepackage{algorithmic}
\usepackage{graphicx}
\usepackage{textcomp}

\usepackage{epstopdf}

\newcommand{\R}{\mathbb{R}}
\renewcommand{\S}{\mathbb{S}}
\newcommand{\N}{\mathbb{N}}
\newcommand{\bmat}[1]{\begin{bmatrix}#1\end{bmatrix}}
\newcommand{\norm}[1]{\left\lVert{#1}\right\rVert}

\newcommand{\diag}{\mathop{\mathrm{diag}}}

\newcommand{\ip}[2]{\left\langle #1, #2 \right\rangle}

\newcommand{\mcl}[1]{\mathcal{ #1}}
\newcommand{\mbf}[1]{\mathbf{ #1}}

\newcommand{\stsp}[4]{\left[\begin{array}{c|c}
      #1 & #2 \\ \hline
      #3 & #4
    \end{array}\right]}

\newtheorem{thm}{Theorem}

\newtheorem{lem}[thm]{Lemma}

\newtheorem{cor}[thm]{Corollary}

\def\QEDclosed{\mbox{\rule[0pt]{1.3ex}{1.3ex}}} 

\def\QED{\QEDclosed} 

\let\bbl\Bigl
\let\bbbl\biggl

\let\bbr\Bigr
\let\bbbr\biggr

\begin{document}

\title{\LARGE \bf
A Convex Solution of the $H_\infty$-Optimal Controller Synthesis Problem for Multi-Delay Systems
}

\author{Matthew~M.~Peet,~\IEEEmembership{Member,~IEEE,}%
\thanks{M. Peet is with the School for the Engineering of Matter, Transport and Energy, Arizona State University, Tempe, AZ, 85298 USA. e-mail: {\tt \small mpeet@asu.edu } } }

\maketitle

\begin{abstract}
Optimal controller synthesis is a bilinear problem and hence difficult to solve in a computationally efficient manner. We are able to resolve this bilinearity for systems with delay by first convexifying the problem in infinite-dimensions - formulating the $H_\infty$ optimal state-feedback controller synthesis problem for distributed-parameter systems as a Linear Operator Inequality - a form of convex optimization with operator variables. Next, we use positive matrices to parameterize positive ``complete quadratic'' operators - allowing the controller synthesis problem to be solved using Semidefinite Programming (SDP). We then use the solution to this SDP to calculate the feedback gains and provide effective methods for real-time implementation. Finally, we use several test cases to verify that the resulting controllers are \textit{optimal} to several decimal places as measured by the minimal achievable closed-loop $H_\infty$ norm, and as compared against controllers designed using high-order Pad\'e approximations.
\end{abstract}
\begin{IEEEkeywords}
Delay Systems, LMIs, Controller Synthesis.\vspace{-2mm}
\end{IEEEkeywords}

\section{Introduction}
To control systems with delay, we must account for the transportation and flow of information. Although solutions to equations of the form\vspace{-1mm}
\[
\dot x(t)=A_0x(t)+A_1x(t-\tau)+Bu(t)\vspace{-1mm}
\]
appear to be functions of time, they are better understood as functions of both time and space:\vspace{-1mm}
\begin{align*}
  \dot x(t) & =Ax(t)+A_1v(t,-\tau)+Bu(t) \\
  \partial_t v(t,s) & =\partial_s v(t,s),\quad v(t,0)=x(t).
\end{align*}
That is, instead of being lost, the state information, $x(t)$, is preserved as $v(t,0)$, transported through a hidden process ($\partial_t v=\partial_s v$), moving at fixed velocity ($-1 m/s$), through a pipe of fixed length ($\tau m$), emerges a fixed time later ($t+\tau$) as $v(t+\tau,-\tau)$, and influences the evolution at that future time ($\dot x(t+\tau)$).

The implication is that feedback controllers for systems with delay must account for both the visible part of the state, $x(t)$, and the hidden process, $v(t,s)$. This concept is well-established and is expressed efficiently in the use of Lyapunov-Krasovskii (LK) functions - a concept dating back to at least 1959~\cite{krasovskii_1963}. LK functionals $V(x,v)$ map $V: \R^n \times L_2^n \rightarrow \R^+$ and offer a method for combining the states, both current ($x$) and hidden ($v$) into a single energy metric.

While the concept of a LK functional may seem obvious, this same logic has been relatively neglected in the design of feedback controllers for time-delay systems. That is, a controller should not only account for the present state, $x(t)\in \R^n$, but should also react to the hidden state $v(t) \in L_2$.

The reason for the relative neglect of the hidden state lies in the development of LMI methods for control in the mid-1990s. Specifically, Ricatti equations and later LMIs were shown to be reliable and practical computational tools for designing optimal and robust controllers for finite-dimensional systems. As a result, research on stability and control of time-delay systems focused on developing clever ways to suppress the infinite-dimensional nature of the hidden state and apply LMIs to a resulting problem in $\R^n$ - for which these tools were originally designed. For example, model transformations were used in~\cite{li_1997,cao_1998,park_1999}, resulting in a Lyapunov function of the form $V(x,v)=z^TMz$ where\vspace{-1mm}
\[
z(t)=x(t-\tau)+\int_{t-\tau}^t \left(A_0 x(s)+A_1 x(s-\tau)\right)ds.\vspace{-1mm}
\]
More recently, Jenson's inequality and free-weighting matrices have been used to parameterize ever more complex Lyapunov functions by projecting the distributed hidden state, $v$ onto a finite-dimensional vector. Indeed, this approach was recently formalized and made infinitely scalable in~\cite{seuret_2014} using a projection-based approach so that for any set of basis functions, $L_i(s)$, we may define an expanded finite-dimensional vector\vspace{-1mm}
\[
z_i(t)=\int_{-\tau}^0 L_i(s)v(t,s)ds.\vspace{-1mm}
\]
so that the resulting Lyapunov function becomes $V(x,v)=z^T M z$ where the size of $M$ increases with the number of basis functions.

Given that LMIs were developed for finite-dimensional systems, the desire to project the hidden state $v\in L_2$, onto a finite-dimensional vector space is  understandable. However, this approach severely limits our ability to perform controller synthesis. Specifically, these projections from $\mcl P: (x,v)\mapsto z$ are not invertible. This is problematic, since standard methods for controller synthesis require the state transformation $\mcl P$ to be invertible - from primal state $(x,v)$ to dual state $(\hat x,\hat v)$). In this approach, the controllers are then designed for the dual state $u(t)=\mcl Z (\hat x, \hat v)$ and then implemented on the original state using the inverse transformation $u(t)=\mcl Z \mcl P^{-1}(x,v)$.

In contrast to projection-based approaches, in this paper and its companion~\cite{peet_2019}, we initially ignore the limitations of the LMI framework and directly formulate convex controller synthesis conditions on an infinite-dimensional space. Specifically, in~\cite{peet_2019}, we formulated convex stabilizing controller synthesis conditions directly in terms of existence of a invertible state transformation $\mcl P:(x,v)\mapsto (\hat x,\hat v)$ and a dual control operator $\mcl Z:(\hat x(t), \hat v(t))\mapsto u(t)$. In Section~\ref{sec:synthesis_DPS}, these results are extended to provide a convex formulation of the $H_\infty$-optimal full-state feedback controller synthesis problem for a general class of Distributed Parameter Systems (DPS).

Having developed a convex formulation of the controller synthesis problem, the question becomes how to test feasibility of these conditions using LMIs - a tool developed for optimization of positive matrix variables (NOT positive operators). As discussed above, a natural approach is to find a way to project these operators onto a finite-dimensional state space (wherein they become matrices) and indeed, one can view the work of~\cite{krasovskii_1962,ross_1969} (or in the PDE case~\cite{lasiecka_1995}) as an attempt to do exactly this. However, these works were unable to recover controller gains and furthermore, the feasibility conditions proposed in~\cite{peet_2019} and in Theorem~\ref{thm:Hinf_DPS} explicitly prohibit such an approach, as they require the positive operator to be coercive and a projected operator will necessarily have a non-trivial null-space.

Because projection is not an option, in this paper and in~\cite{peet_2019}, we have proposed to reverse the dominant paradigm by not \textit{narrowing} the control problem to a finite-dimensional space (where we can apply LMIs), but instead to \textit{expand} the LMI toolset to explicitly allow for parametrization and optimization of operator variables. To understand how this works, let us now discard ODE-based LK functions of the form $V(x,v)=x^T M x$  and instead focus on LK functions of the form
\[
V(x,v):=\int_{-\tau}^0v(s)Mv(s)ds
\]
where the LK function is positive if $M\ge 0$. Now, following the same logic presented above, we increase the complexity of the Lyapunov function by replacing $v(s):s \mapsto \R^n$ with $z(s):s \mapsto \R^q$ defined as
\[
z(s)=\bmat{x\\Z(s)v(s)\\\int_{-\tau}^0Z(s,\theta)v(\theta)d \theta}
\]
where $Z(s)$ and $Z(s,\theta)$ are vectors of functions and increase the dimension of $M$ and hence the complexity of the LK function --- resulting in the well-known class of ``complete-quadratic'' functions. The advantage of this approach, then, is that the resulting LK function can also be represented as
\[
V(x,v):=\int_{-\tau}^0\bmat{x\\v(s)}\left(\mcl P\bmat{x\\v(\cdot)}\right)(s)ds
\]
where
\[
\left(\mcl P\bmat{x\\v}\right)(s)=\bmat{Px + \int_{-\tau}^0Q(\theta)v(\theta)d \theta \\ Q(s)^Tx + S(s)v(s)+\int_{-\tau}^0R(s,\theta)v(\theta)d \theta}
\]
for some $P$, $Q$, $S$ and $R$ (Defined in Theorem~\ref{thm:pos_op_joint}). In this way, positive matrices represent not just positive LK functions (of the complete-quadratic type) but also positive operators in a standardized form - denoted $\mcl P_{\{P,Q,S,R\}}$. This means that if we assume our operators to have this standard form, we can enforce positivity using LMI constraints. Furthermore, linear constraints on the matrix $P$ and the functions $Q$, $R$ and $S$ translate to linear constraints on the elements of the positive matrix $M$.

The contribution of Section~\ref{sec:synthesis_MDS}, then, is to assume all operators have the $PQRS$ form and state conditions on the functions $P$, $Q$, $R$ and $S$ such that the resulting operators satisfy the conditions of Theorem~\ref{thm:Hinf_DPS}. Positivity is then formulated as an LMI constraint in Section~\ref{sec:synthesis_LMI}.

One of the drawbacks of the proposed approach is that the resulting controllers are expressed as operators - of the form $u(t)=\mcl Z \mcl P_{\{P,Q,S,R\}}^{-1}(x(t),v(t))$. The solution to the LMI yields numerical values of operator $\mcl Z$ and functions $P$, $Q$, $R$ and $S$. However, in order to compute the controller gains,
\[
u(t)=K_1 x(t)+K_2v(t,-\tau)+\int_{-\tau}^0K_3(s)v(t,s)ds
\]
we need to find $\hat P$, $\hat Q$, $\hat R$ and $\hat S$ such that $\mcl P_{\{\hat P,\hat Q,\hat S,\hat R\}}=\mcl P_{\{P,Q,S,R\}}^{-1}$. This problem is solved in Section~\ref{sec:inverse} (which is a generalization of the result in~\cite{miao_2017IFAC}) by derivation of an analytic expression for $\hat P$, $\hat Q$, $\hat R$ and $\hat S$ in terms of $P$, $ Q$, $ R$ and $ S$. Finally, practical implementation requires an efficient numerical scheme for calculating $u(t)$ in real-time. This issue is resolved in Section~\ref{sec:controller_construction}.

To make the results of this paper more broadly useful, we have developed efficient implementations for: solving the LMI; calculating the feedback gains; and simulating the closed-loop response. These are available online via Code Ocean and at~\cite{mmpeet_web}. In Section~\ref{sec:validation}, the results are shown to be non-conservative to several decimal places by calculating the minimal achievable closed-loop $H_\infty$-norm bound  for several systems and comparing to results obtained using high-order Pad\'e approximations of the same systems. Obviously, these results presented in this paper are significantly better than any known algorithm for controller synthesis with provable performance metrics. Furthermore, these result can be extended in obvious ways to robust control with uncertainty in system parameters or in delay.

As a final note, the reader should be aware that although the discussion here is for a single delay, the results developed are for multiple delays - a case which requires additional mathematical formalism.

\subsection{Notation}\label{sec:Notation}
Shorthand notation used throughout this paper includes the Hilbert spaces $L_2^m[X]:=L_2(X;\R^m)$ of square integrable functions from $X$ to $\R^m$  and $W^m_2[X]:=W^{1,2}(X;\R^m)=H^{1}(X;\R^m)=\{x\, :\, x, \dot x \in L_2^m[X]\}$. We use $L_2^m$ and $W_2^m$ when domains are clear from context. We also use the extensions $L_2^{n \times m}[X]:=L_2(X;\R^{n \times m})$ and $W_2^{n \times m}[X]:=W^{1,2}(X;\R^{n\times m})$ for matrix-valued functions. $S^n \subset \R^{n \times n}$ denotes the symmetric matrices. We say an operator $\mathcal{P}:Z \rightarrow Z$ is positive on a subset $X$ of Hilbert space $Z$ if $\ip{x}{\mathcal{P}x}_Z\ge 0$ for all $x \in X$. $\mathcal{P}$ is coercive on $X$ if $\ip{x}{\mathcal{P}x}_Z\ge \epsilon \norm{x}_Z^2$ for some $\epsilon>0$ and for all $x \in X$. Given an operator $\mathcal{P}:Z \rightarrow Z$ and a set $X\subset Z$, we use the shorthand $\mcl P(X)$ to denote the image of $\mcl P$ on subset $X$. $I_n \in \S^n$ denotes the identity matrix. $0_{n\times m} \in \R^{n\times m}$ is the matrix of zeros with shorthand $0_{n}:=0_{n \times n}$. We will occasionally denote the intervals $T_i^{j}:=[-\tau_{i},-\tau_{j}]$ and $T_i^{0}:=[-\tau_{i},0]$. For a natural number, $K \in \N$, we adopt the index shorthand notation which denotes $[K]=\{1,\cdots,K\}$. The symmetric completion of a matrix is denoted $*^T$.

%
%
%
%
%
%




\section{The LMI for $H_\infty$-Optimal Controller Synthesis for ODEs }\label{sec:synthesis_ODE}
To better understand the derivation of the main result in Theorem~\ref{thm:Hinf_DPS}, it is instructive to examine the same result in finite dimensions. This is because much of the proof of Theorem~\ref{thm:Hinf_DPS} is a simple generalization of the proof of the ODE synthesis LMI (Lemma~\ref{lem:synthesis_ODE}). Indeed, it is important to state that one of the advantages of the $PQRS$ framework (described above and in Section~\ref{sec:SOS}) is that it \textit{simplifies} the process of controlling systems with delay. Indeed, equipped with this framework and with the theoretical justification provided in~\cite{peet_2019}, almost any LMI developed for estimation and control of ODEs may be generalized and solved for delay systems (using the highly optimized DelayTOOLS extension to SOSTOOLS~\cite{prajna_2002}). To illustrate, consider the ODE system
\begin{align*}
\dot{x}(t)&=Ax(t)+B_1 w(t)+B_2 u(t),\quad x(0)=0\\
  y(t)&=Cx(t)+D_1w(t)+D_2u(t).
\end{align*}

Then the following LMI provides a necessary and sufficient condition for existence of an $H_\infty$-optimal full-state feedback controller.

\begin{lem}[Full-State Feedback Controller Synthesis] \label{lem:synthesis_ODE}
Define:
\[
\hat G(s) = \stsp{A+B_2K}{B_1}{C+D_{2}K}{D_{1}}.
\]
The following are equivalent.
\begin{itemize}
\item There exists a $K$ such that $\displaystyle \norm{\hat G}_{H_\infty}< \gamma$.
\item There exists a $P>0$ and $Z$ such that
\end{itemize}
{\small
\[
\bmat{P A^T  + AP +Z^T B_2^T + B_2 Z & B_1 & P C_1^T + Z^T D_{2}^T\\B_1^T  & -\gamma I&D_{1}^T\\C_1P +D_{2}Z&D_{1}& -\gamma I}<0
\]
}
\end{lem}
\begin{proof}
The proof is a straightforward application of the KYP lemma, a duality transformation, and the Schur Complement lemma. However, since the purpose of this proof is to motivate the proof of Theorem~\ref{thm:Hinf_DPS}, we do not rely on these classical results and instead prove the lemma based on first-principles. In addition, we only prove sufficiency of the non-strict inequality since the necessity proof of the KYP lemma does not easily generalize. First define the storage function $V(x)=x^T P^{-1}x$. Let $u(t)=ZP^{-1}x(t)$. Then if $x(t)$ is a solution of system $\hat G$,
  \begin{align*}
    & \dot V(t)  =x(t)^T P^{-1} ( A x(t)+ B_2 Z P^{-1} x(t)+  B_1 w(t)) + ( A x(t)+ B_2 Z P^{-1} x(t)+  B_1 w(t))^T P^{-1} x(t)\\
    &=\bmat{x(t)\\w(t)}^T \bmat{P^{-1} (A + B_2 Z P^{-1}) + *^T & *^T \\
     B_1^T P^{-1}& 0}\bmat{x(t)\\w(t)}\\
    &=\bmat{P^{-1}x(t)\\w(t)}^T \bmat{ A P + B_2 Z  + *^T & *^T \\
     B_1^T & 0}\bmat{P^{-1}x(t)\\w(t)}
  \end{align*}
Now let $z(t)=P^{-1}x(t)$. Then for any $v$, the matrix inequality implies
\[
\bmat{z\\w\\v}^T\bmat{AP + B_2 Z +*^T & *^T & *^T \\B_1^T  & -\gamma I&*^T\\C_1P +D_{2}Z&D_{1}& -\gamma I}\bmat{z\\w\\v}\le 0
\]
Following the proof of the Schur complement lemma, we let $v=\frac{1}{\gamma}((C_1P +D_{2}Z)z+D_{1}w)$, which implies
\begin{align*}
&\bmat{z\\w\\v}^T\bmat{AP + B_2 Z +*^T & *^T & *^T \\B_1^T  & -\gamma I&*^T\\C_1P +D_{2}Z&D_{1}& -\gamma I}\bmat{z\\w\\v}\\
&=\bmat{z\\w}^T\bmat{AP + B_2 Z +*^T & *^T  \\B_1^T  & -\gamma I}\bmat{z\\w}+\frac{1}{\gamma}\bmat{z\\w}^T\bmat{(C_1P +D_{2}Z)^T\\D_{1}^T}\bmat{C_1P +D_{2}Z&D_{1}}\bmat{z\\w}\le0
\end{align*}
Applying this to $\dot V$, we find{\small
  \begin{align*}
    & \dot V(t)  =\bmat{P^{-1}x(t)\\w(t)}^T \bmat{ A P + B_2 Z  + *^T & *^T \\
     B_1^T & 0}\bmat{P^{-1}x(t)\\w(t)}\\
     &\le \gamma\norm{w}^2-\frac{1}{\gamma}\bmat{z\\w}^T\bmat{(C_1P +D_{2}Z)^T\\D_{1}^T}\bmat{C_1P +D_{2}Z&D_{1}}\bmat{z\\w}\\
     &= \gamma\norm{w}^2-\frac{1}{\gamma}\norm{(C_1P +D_{2}Z)z(t)+D_{1}w(t)}^2\\
     &= \gamma\norm{w}^2-\frac{1}{\gamma}\norm{C_1x(t) +D_{2}Kx(t)+D_{1}w(t)}^2\\
     &= \gamma\norm{w}^2-\frac{1}{\gamma}\norm{y(t)}^2.
  \end{align*}}
  If $w=0$, the LMI implies the system is exponentially stable, which implies $\lim_{t\rightarrow \infty}\norm{x(t)}=0$, which implies  $\lim_{t\rightarrow \infty}V(t)=0$. Since $V(0)=0$, integrating the inequality forward in time, we obtain
\[
\frac{1}{\gamma}\norm{y}_{L_2}^2\le \gamma \norm{w}_{L_2}^2
\]
which completes the proof.
\end{proof}
In the following section, we replicate these steps, simply expanding
\[
\bmat{z\\w\\v}^T\bmat{AP + B_2 Z +*^T & *^T & *^T \\B_1^T  & -\gamma I&*^T\\C_1P +D_{2}Z&D_{1}& -\gamma I}\bmat{z\\w\\v}<0
\]
and replacing terms such as $z^TAPz$ with inner products on the appropriate function space as in $\ip{\mbf z}{\mcl A\mcl P \mbf z}$. Here the bold version of $\mbf z$ emphasizes this term lies in a function space and the calligraphic notation $\mcl A$ indicates $\mcl A$ is an operator.

\section{An Convex Formulation of the Controller Synthesis Problem for Distributed Parameter Systems}\label{sec:synthesis_DPS}

Consider the generic distributed-parameter system
\begin{align}
\dot{\mbf x}(t) &=\mcl A \mbf x(t)+\mcl B_1 w(t) + \mcl B_2 u(t),\quad \mbf x(0)=0, \notag \\
y(t)&=\mcl C \mbf x(t)+D_1 w(t) + \mcl D_2 u(t),\label{eqn:DPS}
\end{align}
where $\mcl A:X \rightarrow Z$, $\mcl B_1: \R^m \rightarrow Z$, $\mcl B_2:U \rightarrow Z$, $\mcl C : X \rightarrow \R^q$, $D_1: \R^m \rightarrow \R^q$, and $\mcl D_2:U \rightarrow \R^q$.

We begin with the following mathematical result on duality, which is a reduced version of Theorem~3 in~\cite{peet_2019}.

\begin{thm} \label{thm:dual}
Suppose $\mcl P$ is a bounded, coercive linear operator $\mathcal{P} : X \rightarrow X$ with $\mathcal{P}(X)=X$ and which is self-adjoint with respect to the $Z$ inner product. Then $\mathcal{P}^{-1}$: exists; is bounded; is self-adjoint; $\mathcal{P}^{-1}: X \rightarrow X$; and $\mathcal{P}^{-1}$ is coercive.
\end{thm}

Using Theorem~\ref{thm:dual}, we give a convex formulation of the $H_\infty$ optimal full-state feedback controller synthesis problem. This result combines: a) a relatively simple extension of the Schur complement Lemma to infinite dimensions; with b) the dual synthesis condition in~\cite{peet_2019}. We note that the ODE equivalent (Lemma~\ref{lem:synthesis_ODE})of this theorem is necessary and sufficient and the proof structure can be credited with, e.g.~\cite{bernussou_1989}.

\begin{thm}\label{thm:Hinf_DPS}
Suppose there exists an $\epsilon>0$, an operator $\mcl P:Z\rightarrow Z$ which satisfies the conditions of Theorem~\ref{thm:dual}, and an operator $\mcl Z: X \rightarrow U$ such that
\begin{align*}
  & \ip{ \mcl A  \mcl P \mbf z}{\mbf z}_Z+\ip{\mbf z}{ \mcl A  \mcl P \mbf z}_Z+\ip{ \mcl B_2  \mcl Z \mbf z}{\mbf z}_Z+\ip{\mbf z}{ \mcl B_2  \mcl Z \mbf z}_Z\\
  & + \ip{\mbf z}{\mcl B_1 w}_Z+ \ip{\mcl B_1 w}{\mbf z}_Z\le \gamma w^Tw - v^T(\mcl{C}\mcl{P}\mbf z)-(\mcl{C} \mcl P \mbf z)^Tv\\
  &-v^T(\mcl{D}_2\mcl{Z}\mbf z) - (\mcl{D}_2 \mcl Z \mbf z)^Tv - v^T(D_1 w)- (D_1w)^Tv\\
  & + \gamma v^Tv-\epsilon \norm{z}_Z^2
\end{align*}
for all $\mbf z \in X$, $w \in \R^m$, and $v \in \R^q$. Then for any $w \in L_2$, if $\mbf x(t)$ and $y(t)$ satisfy $\mbf x(t) \in X$ and
\begin{align}
\dot{\mbf x}(t) &=(\mcl A + \mcl B_2 \mcl Z \mcl P^{-1}) \mbf x(t)+\mcl B_1 w(t) \notag \\
y(t)&=(\mcl C +\mcl D_2 \mcl Z \mcl P^{-1}) \mbf x(t)+D_1 w(t)\label{eqn:DPS_CL}
\end{align}
for all $t \ge 0$, then $\norm{y}_{L_2}\le \gamma \norm{w}_{L_2}$.
\end{thm}

\begin{proof}
By Theorem~\ref{thm:dual} $\mathcal{P}^{-1}$: exists; is bounded; is self-adjoint; $\mathcal{P}^{-1}: X \rightarrow X$; and is coercive.

For $w \in L_2$, let $\mbf x(t)$ and $y(t)$ be a solution of
\begin{align*}
\dot{\mbf x}(t) &=(\mcl A + \mcl B_2 \mcl Z \mcl P^{-1}) \mbf x(t)+\mcl B_1 w(t) \notag \\
y(t)&=(\mcl C +\mcl D_2 \mcl Z \mcl P^{-1}) \mbf x(t)+D_1 w(t)
\end{align*}
such that $\mbf x(t) \in X$ for any finite $t$.

Define the storage function $V(t)=\ip{\mbf x(t)}{\mcl P^{-1} \mbf x(t)}_Z$. Then $V(t)\ge \delta \norm{\mbf x(t)}_Z^2$ for some $\delta>0$. Define $\mbf z(t)=\mcl P^{-1}\mbf x(t)\in X$. Differentiating the storage function in time, we obtain{
  \begin{align*}
    & \dot V(t)  =\ip{\mbf x(t)}{\mcl P^{-1} (\mcl A \mbf x(t)+\mcl B_2 \mcl Z \mcl P^{-1}\mbf x(t)+ \mcl B_1 w(t))}_Z +\ip{\mcl P^{-1} (\mcl A \mbf x(t)+\mcl B_2 \mcl Z \mcl P^{-1}\mbf x(t) + \mcl B_1 w(t))}{\mbf x(t)}_Z\\
    &=\ip{\mcl P^{-1} \mbf x(t)}{ \mcl A \mbf x(t)}_Z+\ip{\mcl P^{-1}\mbf x(t)}{ \mcl B_2 \mcl Z \mcl P^{-1}\mbf x(t) }_Z+\ip{\mcl P^{-1} \mbf x(t)}{ \mcl B_1 w(t)}_Z +\ip{ \mcl A \mbf x(t)}{\mcl P^{-1}\mbf x(t)}_Z\\
    &\quad+\ip{ \mcl B_2 \mcl Z \mcl P^{-1}\mbf x(t) }{\mcl P^{-1}\mbf x(t)}_Z+\ip{ \mcl B_1 w(t)}{\mcl P^{-1} \mbf x(t)}_Z\\
    &=\ip{\mbf z(t)}{ \mcl A \mcl P \mbf z(t)}_Z+\ip{ \mcl B_2 \mcl Z \mbf z(t)}{\mbf z(t)}_Z+\ip{ \mbf z(t)}{ \mcl B_1 w(t)}_Z+\ip{ \mcl A \mcl P \mbf z(t)}{\mbf z(t)}_Z+\ip{\mbf z(t)}{ \mcl B_2 \mcl Z \mbf z(t) }_Z+\ip{ \mcl B_1 w(t)}{ \mbf z(t)}_Z\\
&  \le \gamma w(t)^Tw(t) -v(t)^T(\mcl{CP}\mbf z(t))-(\mcl{CP}\mbf z(t))^Tv(t)-v(t)^T(\mcl{D}_2\mcl{Z}\mbf z(t))-(\mcl{D}_2 \mcl Z \mbf z(t))^Tv(t) - v(t)^T(D_1w(t))\\
&\quad- (D_1w(t))^Tv(t) + \gamma v(t)^Tv(t)-\epsilon \norm{\mbf z(t)}_Z^2\\
&  = \gamma w(t)^Tw(t) -v(t)^T\left((\mcl{C} + \mcl D_2 \mcl Z \mcl P^{-1})\mbf x(t)+D_1w(t)\right)-\left((\mcl{C} + \mcl D_2 \mcl Z \mcl P^{-1})\mbf x(t)+D_1w(t)\right)^Tv(t)  + \gamma v(t)^Tv(t)-\epsilon \norm{\mbf z(t)}_Z^2\\
&  = \gamma w(t)^Tw(t) -v(t)^Ty(t)-y(t)^Tv(t)  + \gamma v(t)^Tv(t)-\epsilon\norm{\mbf z(t)}_Z^2
  \end{align*}}
  for any $v(t) \in \R^q$ and all $t \ge 0$. Choose $v(t) = \frac{1}{\gamma}y(t)$ and we get
  \begin{align*}
    \dot V(t) & \le \gamma \norm{w(t)}^2 -\frac{2}{\gamma}\norm{y(t)}^2  + \frac{1}{\gamma}\norm{y(t)}^2-\epsilon \norm{\mbf z(t)}_Z^2\\
    & = \gamma \norm{w(t)}^2 -\frac{1}{\gamma}\norm{y(t)}^2-\epsilon \norm{\mbf z(t)}_Z^2.
  \end{align*}
  Since $\mcl P$ is bounded, there exists a $\sigma>0$ such that
\[
V(t)=\ip{\mbf x(t)}{\mcl P^{-1}\mbf x(t)}_Z=\ip{\mbf z(t)}{\mcl P\mbf z(t)}_Z\le \sigma \norm{\mbf z(t)}_Z^2.
\]
We conclude, therefore, that
\[
\dot V(t) \le -\frac{\epsilon}{\sigma} V(t)+\gamma \norm{w(t)}^2-\frac{1}{\gamma}\norm{y(t)}^2.
\]
Therefore, since $w \in L_2$, we may conclude by Gronwall-Bellman that $\lim_{t\rightarrow \infty} V(t)=0$. Integrating this expression forward in time, and using $V(0)=V(\infty)=0$, we obtain
\[
\frac{1}{\gamma}\norm{y}_{L_2}^2 \le \gamma \norm{w}_{L_2}^2
\]
which concludes the proof.

\end{proof}


\section{Theorem~\ref{thm:Hinf_DPS} Applied to Multi-Delay Systems} \label{sec:synthesis_MDS}
Theorem~\ref{thm:Hinf_DPS} gives a convex formulation of the controller synthesis problem for a general class of distributed-parameter systems. In this section and the next, we apply Theorem~\ref{thm:Hinf_DPS} to the case of systems with multiple delays. Specifically, we consider solutions to the system of equations given by
\begin{align}
\dot x(t)&=A_0x(t)+\sum_i A_ix(t-\tau_i)+B_1w(t)+B_2u(t)\notag \\
y(t)&=C_0x(t)+\sum_i C_i x(t-\tau_i)+D_1w(t) + D_2u(t)\label{eqn:MDS}
\end{align}
where $w(t)\in \R^{m}$ is the disturbance input, $u(t)\in \R^{p}$ is the controlled input,  $y(t) \in \R^q$ is the regulated output, $x(t)$ are the state variables and $\tau_i >0$ for $i\in [1,\cdots,K]$ are the delays ordered by increasing magnitude. We assume $x(s)=0$ for $s \in [-\tau_K,0]$.

Our first step, then, is to express System~\eqref{eqn:MDS} in the abstract form of~\eqref{eqn:DPS}. Following the mathematical formalism developed in~\cite{peet_2019}, we define the inner-product space
 $Z_{m,n,K}:=\{\R^m \times L_2^{n}[-\tau_1,0]\times \cdots \times L_2^n[-\tau_K,0]\}$ and for $\{x,\phi_1,\cdots,\phi_K\}\in Z_{m,n,K}$, we define the following shorthand notation
\[
\bmat{x\\ \phi_i}:=\{x,\phi_1,\cdots,\phi_K\},
\]
which allows us to simplify expression of the inner product on $Z_{m,n,K}$, which we define to be
\[
\ip{\bmat{y\\ \psi_i}}{\bmat{x\\ \phi_i}}_{Z_{m,n,K}}=\tau_K y^T x + \sum_{i=1}^K \int_{-\tau_i}^0 \psi_i(s)^T\phi_i(s)ds.
\]
When $m=n$, we simplify the notation using $Z_{n,K}:=Z_{n,n,K}$. The state-space for System~\eqref{eqn:MDS} is defined as
\[
X:=\left\{\bmat{x \\ \phi_i} \in Z_{n,K}\, : \, \substack{ \phi_i \in W_2^n[-\tau_i,0] \text{ and }\\ \phi_i(0)=x \text{ for all } i\in [K]} \right\}.
\]
Note that $X$ is a subspace of $Z_{n,K}$ and inherits the norm of $Z_{n,K}$. We furthermore extend this notation to say
\[
\bmat{x \\ \phi_i}(s)=\bmat{y \\ f(s,i)}
\]
if  $x=y$ and $\phi_i(s)=f(s,i)$ for $s \in [-\tau_i,0]$ and $i\in [K]$.

We now represent the infinitesimal generator, $\mathcal{A}:X\rightarrow Z_{n,K}$, of Eqn.~\eqref{eqn:MDS} as
\[
\mathcal{A}\bmat{x\\ \phi_i}(s):= \bmat{A_0 x + \sum_{i=1}^K A_i \phi_i(-\tau_i)\vspace{2mm}\\ \dot \phi_i(s)}.
\]
Furthermore, $\mcl B_1: \R^m \rightarrow Z_{n,K}$, $\mcl B_2: \R^p \rightarrow Z_{n,K}$, $\mcl D_1: \R^m \rightarrow \R^q$, $\mcl D_2: \R^p \rightarrow \R^q$, and $\mcl C : Z_{n,K} \rightarrow \R^p$ are defined as
\begin{eqnarray*}
 &(\mathcal{B}_1w)(s):=\bmat{B_1 w\\0},\; \quad (\mathcal{B}_2u)(s):=\bmat{B_2 u\\0},\\
  &\quad \left(\mcl C \bmat{\psi\\ \phi_i}\right):=\bmat{C_0 \psi + \sum_i C_i \phi_i(-\tau_i)},\; \quad  \\
  &(\mathcal{D}_1w)(s):=\bmat{D_1 w}, \quad (\mathcal{D}_2u)(s):=\bmat{D_2 u}\\
\end{eqnarray*}
Having defined these operators, we note that for any solution $x(t)$ of Eqn.~\eqref{eqn:MDS}, using the above notation if we define
\[
\left(\mbf x(t)\right)(s)=\bmat{\mbf x_1(t) \\ \mbf x_{2}(t)}(s)=\bmat{x(t)\\x(t+s)}
\]
Then $\mbf x$ satisfies Eqn.~\eqref{eqn:DPS} using the operator definitions given above. The converse statement is also true.

\subsection{A Parametrization of Operators}\label{subsec_PQRS}

We now introduce a class of operators $\mathcal{P}_{\{P,Q_i,S_i,R_{ij}\}}:Z_{m,n,K}\rightarrow Z_{m,n,K}$, parameterized by matrix $P$ and matrix-valued functions $Q_i\in W_2^{m\times n}[-\tau_i,0]$, $S_i\in W_2^{n\times n}[-\tau_i,0]$, $R_{ij}\in W_2^{n\times n}\left[[-\tau_i,0]\times[-\tau_j,0]\right]$ as
\begin{align}
&\left(\mathcal{P}_{\{P,Q_i,S_i,R_{ij}\}} \bmat{x\\ \phi_i}\right)(s) :=\bmat{  P x +  \sum_{i=1}^K \int_{-\tau_i}^0 Q_i(s)\phi_i(s) d s \\
\tau_K Q_i(s)^T x\hspace{-.5mm} +\hspace{-.5mm} \tau_K S_i(s)\phi_i(s) \hspace{-.5mm}+ \hspace{-.5mm}\sum_{j=1}^K \int_{-\tau_j}^0 \hspace{-1.5mm} R_{ij}(s,\theta)\phi_j(\theta)\, d \theta. } \notag \\[-6mm]\notag
\end{align}

For this class of operators, the following Lemma combines Lemmas~3 and~4 in~\cite{peet_2019} and gives conditions under which $\mcl{P}_{\{P,Q_i,S_i,R_{ij}\}}$ satisfies the conditions of Theorem~\ref{thm:dual}.

\begin{lem}\label{lem:selfadjoint_MD}
Suppose that $S_i\in W_2^{n\times n}[-\tau_i,0]$, $R_{ij}\in W_2^{n\times n}\left[[-\tau_i,0]\times[-\tau_j,0]\right]$ and $S_i(s)=S_i(s)^T$, $R_{ij}(s,\theta)=R_{ji}(\theta,s)^T$, $P=\tau_KQ_i(0)^T + \tau_KS_i(0)$ and $Q_j(s)=R_{ij}(0,s)$ for all $i,j\in [K]$. Further suppose $\mathcal{P}_{\{P,Q_i,S_i,R_{ij}\}}$ is coercive on $Z_{n,K}$.  Then $\mcl{P}_{\{P,Q_i,S_i,R_{ij}\}}$: is a self-adjoint bounded linear operator with respect to the inner product defined on $Z_{n,K}$;  maps $\mathcal{P}_{\{P,Q_i,S_i,R_{ij}\}}: X \rightarrow X$; and $\mcl{P}_{\{P,Q_i,S_i,R_{ij}\}}(X)=X$.
\end{lem}

Starting in Section~\ref{sec:SOS}, we will assume $Q_i$, $S_i$, and $R_{ij}$ are polynomial and give LMI conditions for positivity of operators of the form $\mathcal{P}_{\{P,Q_i,S_i,R_{ij}\}}$.

\subsection{The Controller Synthesis Problem for Systems with Delay}
Theorem~\ref{thm:Hinf_DPS} gives a convex formulation of the controller synthesis problem, where the data is the $6$ operators $\mcl A$, $\mcl B_1$, $\mcl B_2$, $\mcl C$, $\mcl D_1$, and $\mcl D_2$ and the variables are the operators $\mcl P$ and $\mcl Z$. For multi-delay systems, we have defined the 6 operators and parameterized the decision variables $\mcl P$ using $\mathcal{P}_{\{P,Q_i,S_i,R_{ij}\}}$. We now likewise parameterize the decision variables $\mcl Z:Z_{n,k} \rightarrow \R^p$ using matrices $Z_0$, $Z_{1i}$ and functions $Z_{2i}$ as{\small
\[
\left(\mcl Z \bmat{\psi\\ \phi_i}\right):=\bmat{Z_0 \psi + \sum_i Z_{1i} \phi_i(-\tau_i)+ \sum_i\int_{-\tau_i}^0 Z_{2i}(s) \phi_i(s) ds}.
\]}
The following theorem gives convex constraints on the variables $P$, $Q_i$, $S_i$, $R_{ij}$, $Z_0$, $Z_{1i}$ and $Z_{2i}$  under which Theorem~\ref{thm:Hinf_DPS} is satisfied when $\mcl A$, $\mcl B_1$, $\mcl B_2$, $\mcl C$, $\mcl D_1$, and $\mcl D_2$ are as defined above.

\begin{thm}\label{thm:synthesis_MD}
Suppose that there exist $S_i\in W_2^{n\times n}[-\tau_i,0]$, $R_{ij}\in W_2^{n\times n}\left[[-\tau_i,0]\times[-\tau_j,0]\right]$ and $S_i(s)=S_i(s)^T$ such that $R_{ij}(s,\theta)=R_{ji}(\theta,s)^T$, $P=\tau_KQ_i(0)^T + \tau_KS_i(0)$ and $Q_j(s)=R_{ij}(0,s)$ for all $i,j\in [K]$, and matrices $Z_0 \in \R^{p \times n}$, $Z_{1i}\in \R^{p \times n}$ and $Z_{2i} \in W^{p\times n}_2[T_i^0]$ such that $\ip{\mbf x}{\mathcal{P}_{\{P,Q_i,S_i,R_{ij}\}} \mbf x}_{Z_{n,K}} \ge \epsilon \norm{\mbf x}^2$ for all $\mbf x \in Z_{n,K}$ and\vspace{-2mm}
\[
\ip{\bmat{\bmat{v \\ w \\ y_1 \\ y_2} \\ \phi_i}}{ \mathcal{P}_{\{D,E_i,\dot S_i,G_{ij}\}}\bmat{\bmat{v \\ w \\ y_1 \\ y_2}\\ \phi_i}}_{Z_{q+m+{n(K+1)},n,K}}\hspace{-1cm}\hspace{-6mm}\le- \epsilon \norm{\bmat{y_1 \\ \phi_i}}_{Z_{n,K}}^2\vspace{-2mm}
\] for all $y_1 \in \R^n$ and $\bmat{\bmat{ v\\ w\\ y_1\\y_2}\\ \phi_i} \in Z_{q+m+n(K+1),n,K}$ where\vspace{-2mm}

{\small
\begin{align*}
&L_{0}:= A_0 P+ \sum_{i=1}^K \left(  \tau_K A_i Q_i(-\tau_i)^T  + \frac{1}{2}S_i(0)\right)+B_2Z_0,   \notag\\
&L_1:=\frac{1}{\tau_K}C_0P+\sum_i C_i Q_i(-\tau_i)^T+\frac{1}{\tau_K}D_2Z_0\\
&L_{2i}:=C_i S_i(-\tau_i)+\frac{1}{\tau_K}D_2Z_{1i}\\
&L_{3i}:=\tau_K A_iS_i(-\tau_i)+B_2Z_{1i}\\
&D=\bmat{
-\frac{\gamma}{\tau_K} I &\frac{1}{\tau_K}D_1&L_1 &L_{21} &\hdots &L_{2K}\\
  *^T&-\frac{\gamma}{\tau_K} I &B_1^T&0&\hdots&0\\
  *^T&*^T&L_0+L_0^T&L_{31} &\hdots &L_{3K}\\
  *^T&*^T&*^T&-S_1(-\tau_1)&\hdots &0\\
  \vdots &\vdots&\vdots&\vdots&\ddots &\vdots\\
  *^T&*^T&*^T&*^T&\hdots &-S_k(-\tau_K)}\\
&E_i(s)= \frac{1}{\tau_K}\cdot\bmat{C_0Q_i(s)+\sum_j C_j R_{ji}(-\tau_j,s)+D_2Z_{2i}(s)\\0\\ \tau_K\left(A_0 Q_i(s) +\dot Q_i(s)+\sum_{j=1}^K A_j R_{ji}(-\tau_j,s)+B_2Z_{2i}(s)\right)  \\ 0 \\ \vdots \\0}\\
&G_{ij}(s,\theta):=\frac{\partial}{\partial s}R_{ij}(s,\theta)+\frac{\partial}{\partial \theta}R_{ji}(s,\theta)^T, \quad i,j \in [K].
\end{align*}
}
Then if
\[
u(t)=\mcl Z \mcl P_{\{P,Q_i,S_i,R_{ij}\}}^{-1}\bmat{x(t)\\x(t+s)}
\]
where{\small
\[
\left(\mcl Z \bmat{x \\ \phi_i}\right)(s):=Z_0 x + \sum_{i=1}^K Z_{1i} \phi_i(-\tau_i) + \sum_{i=1}^K \int_{-\tau_i}^0 \hspace{-3mm}Z_{2i}(s)\phi_i(s)ds,\vspace{-2mm}
\]}
then for any $w \in L_2$, if  $ x(t)$ and $y(t)$ satisfy Eqn.~\eqref{eqn:MDS}, $\norm{y}_{L_2}\le \gamma \norm{w}_{L_2}$.

\end{thm}

\begin{proof}
For any $w\in L_2$, using the definitions of $u(t)$, and $\mcl A$, $\mcl B_1$, $\mcl B_2$, $\mcl C$, $\mcl D_1$, $\mcl D_2$ and $\mcl Z$ given above, $y(t)$ and $x(t)$ satisfy Eqn.~\eqref{eqn:MDS} if and only if $y(t)$ and $\mbf x(t):=\bmat{x(t)\\x(t+s)}$ satisfy Eqn.~\eqref{eqn:DPS}. Therefore, $\norm{y}_{L_2}\le \gamma \norm{w}_{L_2}$ if
\begin{align*}
  & \ip{ \mcl A  \mcl P \mbf z}{\mbf z}_Z+\ip{\mbf z}{ \mcl A  \mcl P \mbf z}_Z+\ip{ \mcl B_2  \mcl Z \mbf z}{\mbf z}_Z+\ip{\mbf z}{ \mcl B_2  \mcl Z \mbf z}_Z+ \ip{\mbf z}{\mcl B_1 w}_Z+ \ip{\mcl B_1 w}{\mbf z}_Z\\
  &\le \gamma w^Tw -v^T(\mcl{C}\mcl{P}\mbf z)-(\mcl{C} \mcl P \mbf z^Tv-v^T(\mcl{D}_2\mcl{Z}\mbf z)-(\mcl{D}_2 \mcl Z \mbf z)^Tv - v^T(D_1 w)- (D_1w)^Tv + \gamma v^T v-\epsilon \norm{z}_Z^2
\end{align*}
for all $\mbf z \in X$, $w \in \R^m$, and $v \in \R^q$.
The rest of the proof is lengthy but straightforward. We simply show that if we define
\[
f = \bmat{\mbf z_{2,1}(-\tau_1)^T& \cdots &\mbf z_{2,K}(-\tau_K)^T}^T,
\]
then
\begin{align}
& \ip{ \mcl A  \mcl P \mbf z}{\mbf z}_Z+\ip{\mbf z}{ \mcl A  \mcl P \mbf z}_Z+\ip{ \mcl B_2  \mcl Z \mbf z}{\mbf z}_Z+\ip{\mbf z}{ \mcl B_2  \mcl Z \mbf z}_Z + \ip{\mbf z}{\mcl B_1 w}_Z + \ip{\mcl B_1 w}{\mbf z}_Z\notag\\
& - \gamma w^Tw +v^T(\mcl{C}\mcl{P}\mbf z)+(\mcl{C} \mcl P \mbf z)^Tv+v^T(\mcl{D}_2\mcl{Z}\mbf z)+(\mcl{D}_2 \mcl Z \mbf z)^Tv + v^T(D_1 w) + (D_1w)^Tv - \gamma v^Tv\label{eqn:expansion}\\
&= \ip{\bmat{\bmat{v \\ w \\ \mbf z_1 \\ f} \\ \mbf z_{2i}}}{ \mathcal{P}_{\{D,E_i,\dot S_i,G_{ij}\}}\bmat{\bmat{v \\ w \\ \mbf z_1 \\ f} \\ \mbf z_{2i}}}_{Z_{q+m+{n(K+1)},n,K}}\hspace{-22mm}\le - \epsilon \norm{\bmat{\mbf z_1 \\ \mbf z_{2i}}}_{Z_{n,K}}^2\notag\\
&=- \epsilon \norm{\mbf z}_{Z_{n,K}}^2.\notag
\end{align}
Before we begin, for convenience and efficiency of presentation, we will denote $m_0:=q+m+n(K+1)$ and
\[
h:=\bmat{v^T & w^T & \mbf z_1^T & f^T}^T.
\]
It may also be helpful to note that the quadratic form defined by a $\mcl{P}_{\{D,E_i,F_i,G_{ij}\}}$ operator expands out as
\begin{align}
&\ip{\bmat{h\\ \mbf z_{2i}}}{\mcl{P}_{\{D,E_i,F_i,G_{ij}\}} \bmat{h\\ \mbf z_{2i}}}_{Z_{m_0,n,K}}\notag \\
&=\tau_K h^T D h +  \tau_K \sum_{i=1}^K \int_{-\tau_i}^0 h^T E_i(s)\mbf z_{2i}(s) d s \notag + \tau_K \sum_i \int_{-\tau_i}^0  \mbf z_{2i}(s)^T E_i(s)^T h ds \notag \\
&\quad + \tau_K \sum_i \int_{-\tau_i}^0 \mbf z_{2i}(s)^T F_i(s)\mbf z_{2i}(s)ds  + \sum_{ij} \int_{-\tau_i}^0 \int_{-\tau_j}^0 \mbf z_{2i}(s)^T G_{ij}(s,\theta)\mbf z_{2j}(\theta)\, d \theta ds.\label{eqn:quadratic}
\end{align}
Our task, therefore, is simply to write all the terms we find in~\eqref{eqn:expansion} in the form of Equation~\eqref{eqn:quadratic} for an appropriate choice of matrix $D$ and functions $E_i$, $F_i$, and $G_{ij}$. Fortunately, the most complicated part of this operation has already been completed. Indeed, from Theorem~5 in~\cite{peet_2019}, we have the first two terms can be represented as

\begin{align*}
&\ip{\mathcal{A}\mathcal{P}\mbf z}{\mbf z}_{Z_{n,K}}+\ip{\mbf z}{\mathcal{A}\mathcal{P}\mbf z}_{Z_{n,K}}
=\ip{\bmat{h\\ \mbf z_{2i}}}{\mathcal{D}\bmat{h \\ \mbf z_{2i}}}_{Z_{m_0,n,K}}\hspace{-2mm},
\end{align*}
where $\mcl D:=\mathcal{P}_{\{D_1,E_{1i},\dot S_i,G_{ij}\}}$ (Do not confuse this $D_1$ with the $D_1$ in Eqn.~\eqref{eqn:DPS}) and
\begin{align*}
&D_1:=\bmat{0&0&0&0& \hdots &0\\
0&0&0&0& \hdots &0\\
0&0&C_{0}+C_0^T & C_{1} & \cdots &C_{k} \\0&0&C_{1}^T &-S_1(-\tau_1) & 0&0 \\
\vdots&\vdots&\vdots&0&\ddots&0 \\
0&0&C_{k^T} &0&0&-S_k(-\tau_K)},\\
&C_{0}\hspace{-.5mm}:=\hspace{-.5mm} A_0 P \hspace{-.5mm}+\hspace{-.5mm}\tau_K \hspace{-.5mm}\sum_{i=1}^K\hspace{-.5mm} (  A_i Q_i(-\tau_i)^T \hspace{-1mm}  + \hspace{-.5mm}\frac{1}{2}S_i(0)),   \notag\\
&C_{i}:=\tau_K A_iS_i(-\tau_i),\quad i \in [K] \\
&E_{1i}(s):=\bmat{0&0&B_i(s)^T & 0 &\cdots &0}^T, \quad i \in [K]\\
&B_{i}(s):=A_0 Q_i(s) +\dot Q_i(s)+\sum_{j=1}^K A_j R_{ji}(-\tau_j,s), \quad i \in [K] \\
&G_{ij}(s,\theta):=\frac{\partial}{\partial s}R_{ij}(s,\theta)+\frac{\partial}{\partial \theta}R_{ji}(s,\theta)^T, \quad i,j \in [K].
\end{align*}


Having already dealt with the most difficult terms, we now start with the easiest. Recalling that


\begin{eqnarray*}
 &(\mathcal{B}_1w)(s):=\bmat{B_1 w\\0},\; (\mathcal{D}_1w)(s):=\bmat{D_1 w},
\end{eqnarray*}
We have $\ip{\mbf z}{\mcl B_1 w}_Z = \tau_K z_1^T B_1 w$ and hence{\small

\begin{align*}
  & \ip{\mbf z}{\mcl B_1 w}_Z+ \ip{\mcl B_1 w}{\mbf z}_Z  -\gamma w^Tw + v^TD_1 w+ (D_1w)^T v - \gamma v^Tv\\
  &=\tau_K \bmat{v\\w\\ \mbf z_1\\ \mbf z_{21}(-\tau_1) \\ \vdots \\ \mbf z_{2K}(-\tau_K)}^T  \underbrace{\frac{1}{\tau_K}\bmat{-\gamma I &D_1&0&0&\hdots&0\\
  D_1^T&-\gamma I &\tau_K B_1^T&0&\hdots&0\\
  0&\tau_K B_1&0&0&\hdots &0\\
  0&0&0&0&\hdots &0\\
  \vdots &\vdots&\vdots&\vdots&\ddots &\vdots\\
  0&0&0&0&\hdots &0}}_{D_0}  \bmat{v\\w\\ \mbf z_1\\ \mbf z_{21}(-\tau_1) \\ \vdots \\ \mbf z_{2K}(-\tau_K)}\\
  &=\ip{\bmat{h\\ \mbf z_{2i}}}{\mcl{P}_{\{D_0,0,0,0\}} \bmat{h\\ \mbf z_{2i}}}_{Z_{m_0,n,K}}
\end{align*}}


Next, we consider the terms
\begin{align*}
  & v^T(\mcl{C}\mcl{P}\mbf z)+(\mcl{C} \mcl P \mbf z)^Tv
\end{align*}
If we recall that
\begin{eqnarray*}
  \left(\mcl C \bmat{\psi\\ \phi_i}\right):=\bmat{C_0 \psi + \sum_i C_i \phi_i(-\tau_i)},\;
\end{eqnarray*}
then we have the expansion{\small
\begin{align*}
  & 2v^T(\mcl{C}\mcl{P}\mbf z)\\
  &=2v^T\bbl[C_0 ( P \mbf z_1 +  \sum_{i=1}^K \int_{-\tau_i}^0 Q_i(s)\mbf z_{2i}(s) d s) + \sum_i C_i \bbl(\tau_K Q_i(-\tau_i)^T \mbf z_1\hspace{-.5mm} + \tau_K S_i(-\tau_i)\mbf z_{2i}(-\tau_i) \hspace{-.5mm}+ \hspace{-.5mm}\sum_{j=1}^K \int_{-\tau_j}^0 \hspace{-1.5mm} R_{ij}(-\tau_i,\theta)\mbf z_{2j}(\theta)\, d \theta\bbr)\bbr]\\
  &=2v^T\bbbl[ \left( C_0P+\sum_i \tau_K C_i Q_i(-\tau_i)^T  \right)\mbf z_1 \\
  &\qquad+ \tau_K \sum_i  C_i S_i(-\tau_i)\mbf z_{2i}(-\tau_i) +  \sum_{i=1}^K \int_{-\tau_i}^0 \left(C_0Q_i(s)\right)\mbf z_{2i}(s) d s  +  \sum_{i=1}^K \int_{-\tau_i}^0 \sum_j C_j R_{ji}(-\tau_j,s)\mbf z_{2i}(s)\, d s \bbbr]\\
  &=2v^T\hspace{-1mm}\tau_K\bbl[ \hspace{-.5mm}\left( \frac{1}{\tau_K}C_0P\hspace{-.5mm}+\hspace{-1mm}\sum_i C_i Q_i(-\tau_i)^T\hspace{-1mm}  \right)\mbf z_1\hspace{-.5mm} + \hspace{-1mm}\sum_i  C_i S_i(-\tau_i)\mbf z_{2i}(-\tau_i)+\hspace{-1mm}\hspace{-1mm} \frac{1}{\tau_K} \sum_{i=1}^K \int_{-\tau_i}^0 \left(C_0Q_i(s)+\sum_j C_j R_{ji}(-\tau_j,s)\right)\mbf z_{2i}(s) d s \bbr]\\
  &=\tau_K\bmat{v\\w\\ \mbf z_1\\ \mbf z_{21}(-\tau_1) \\ \vdots \\ \mbf z_{2K}(-\tau_K)}^T
  \underbrace{\frac{1}{\tau_K}\bmat{0 &0&C_0P+\sum_i \tau_K C_i Q_i(-\tau_i)^T&\tau_K C_1 S_1(-\tau_1)&\hdots&\tau_K C_K S_K(-\tau_K)\\
  *^T&0 &0&0&\hdots&0\\
  *^T&*^T&0&0&\hdots &0\\
  *^T&*^T&*^T&0&\hdots &0\\
  \vdots &\vdots&\vdots&\vdots&\ddots &\vdots\\
  *^T&*^T&*^T&*^T&\hdots &0}}_{D_2} \bmat{v\\w\\ \mbf z_1\\ \mbf z_{21}(-\tau_1) \\ \vdots \\ \mbf z_{2K}(-\tau_K)}\\
  &\qquad +2\tau_K \sum_{i=1}^K \int_{-\tau_i}^0\bmat{v\\w\\ \mbf z_1\\ \mbf z_{21}(-\tau_1) \\ \vdots \\ \mbf z_{2K}(-\tau_K)}^T \underbrace{\frac{1}{\tau_K}\bmat{C_0Q_i(s)+\sum_j C_j R_{ji}(-\tau_j,s)\\0\\0\\0\\ \vdots \\ 0}}_{E_{2i}(s)}\mbf z_{2i}(s) ds.
\end{align*}}
We therefore conclude that{\small
\[
 v^T(\mcl{C}\mcl{P}\mbf z)+(\mcl{C} \mcl P \mbf z)^Tv=\ip{\bmat{h\\ \mbf z_{2i}}}{\mcl{P}_{\{D_2,E_{2i},0,0\}} \bmat{h\\ \mbf z_{2i}}}_{Z_{m_0,n,K}}\hspace{-2mm}\hspace{-.5mm}.
\]}

We now examine the final set of terms which contain $\mcl Z$.
\begin{align*}
  & \ip{ \mcl B_2  \mcl Z \mbf z}{\mbf z}_Z+\ip{\mbf z}{ \mcl B_2  \mcl Z \mbf z}_Z +v^T(\mcl{D}_2\mcl{Z}\mbf z)+(\mcl{D}_2 \mcl Z \mbf z)^Tv
\end{align*}
If we recall that
\begin{eqnarray*}
 &(\mathcal{B}_2u)(s):=\bmat{B_2 u\\0},(\mathcal{D}_2u)(s):=\bmat{D_2 u}
  \end{eqnarray*}
then we have the expansion
\begin{align*}
  & 2\ip{\mbf z}{ \mcl B_2  \mcl Z \mbf z}_Z+2v^T(\mcl{D}_2\mcl{Z}\mbf z)\\
  &=2\tau_K \mbf z_1^T  \bbl[B_2Z_0 \mbf z_1 + \sum_i B_2Z_{1i} \mbf z_{2i}(-\tau_i)+ \sum_i\int_{-\tau_i}^0 B_2Z_{2i}(s) \mbf z_{2i}(s) ds\bbr]\\
  &+2v^T  \bbl[D_2Z_0 \mbf z_1 + \sum_i D_2Z_{1i} \mbf z_{2i}(-\tau_i)+ \sum_i\int_{-\tau_i}^0 D_2Z_{2i}(s)\mbf  z_{2i}(s) ds\bbr]\\
&=\tau_K\bmat{v\\w\\ \mbf z_1\\ \mbf z_{21}(-\tau_1) \\ \vdots \\ \mbf z_{2K}(-\tau_K)}^T  \underbrace{\bmat{
  0 &*^T &*^T &*^T &\hdots &*^T\\
  0 &0 &*^T&*^T&\hdots&*^T\\
  (\frac{1}{\tau_K}D_2Z_0)^T&0&B_2Z_0+Z_0^TB_2^T&*^T &\hdots &*^T\\
  (\frac{1}{\tau_K}D_2Z_{11})^T&0&(B_2Z_{11})^T&0&\hdots &*^T\\
  \vdots &\vdots&\vdots&\vdots&\ddots &\vdots\\
  (\frac{1}{\tau_K}D_2Z_{1K})^T&0&(B_2Z_{1K})^T&0&\hdots &0}}_{D_3}
  \bmat{v\\w\\ \mbf z_1\\ \mbf z_{21}(-\tau_1) \\ \vdots \\ \mbf z_{2K}(-\tau_K)}\\
  &+2\tau_K \sum_{i=1}^K \int_{-\tau_i}^0\bmat{v(t)\\w(t)\\ \mbf z_1(t)\\ \mbf z_{21}(t,-\tau_1) \\ \vdots \\ \mbf z_{2K}(t,-\tau_K)}^T  \underbrace{\frac{1}{\tau_K}\bmat{D_2Z_{2i}(s) \\ 0 \\   \tau_K B_2Z_{2i}(s)    \\0\\ \vdots \\ 0}}_{E_{3i}(s)}\mbf z_{2i}(s) ds.
  \end{align*}
We therefore conclude that
\begin{align*}
&\ip{ \mcl B_2  \mcl Z \mbf z}{\mbf z}+\ip{\mbf z}{ \mcl B_2  \mcl Z \mbf z} +v^T(\mcl{D}_2\mcl{Z}\mbf z)+(\mcl{D}_2 \mcl Z \mbf z)^Tv\\
&=\ip{\bmat{h\\ \mbf z_{2i}}}{\mcl{P}_{\{D_3,E_{3i},0,0\}} \bmat{h\\ \mbf z_{2i}}}_{Z_{m_0,n,K}}
\end{align*}

%

Summing all the terms we have
\begin{align*}
&D=D_{0}+D_{1}+D_{2}+D_{3}\\
&=\frac{1}{\tau_K}\bmat{-\gamma I &D_1&0&0&\hdots&0\\
  D_1^T&-\gamma I &\tau_K B_1^T&0&\hdots&0\\
  0&\tau_K B_1&0&0&\hdots &0\\
  0&0&0&0&\hdots &0\\
  \vdots &\vdots&\vdots&\vdots&\ddots &\vdots\\
  0&0&0&0&\hdots &0}
  +\bmat{0&0&0&0& \hdots &0\\
0&0&0&0& \hdots &0\\
0&0&C_{0}+C_0^T & C_{1} & \cdots &C_{k} \\0&0&C_{1}^T &-S_1(-\tau_1) & 0&0 \\
\vdots&\vdots&\vdots&0&\ddots&0 \\
0&0&C_{k^T} &0&0&-S_k(-\tau_K)}\\
&  +\frac{1}{\tau_K}\bmat{0 &0&C_0P+\sum_i \tau_K C_i Q_i(-\tau_i)^T&\tau_K C_1 S_1(-\tau_1)&\hdots&\tau_K C_K S_K(-\tau_K)\\
  *^T&0 &0&0&\hdots&0\\
  *^T&*^T&0&0&\hdots &0\\
  *^T&*^T&*^T&0&\hdots &0\\
  \vdots &\vdots&\vdots&\vdots&\ddots &\vdots\\
  *^T&*^T&*^T&*^T&\hdots &0}\\
&  +\bmat{0 &0&\frac{1}{\tau_K}D_2Z_0 &\frac{1}{\tau_K}D_2Z_{11} &\hdots &\frac{1}{\tau_K}D_2Z_{1K}\\
  *^T&0 &0&0&\hdots&0\\
  *^T&*^T&B_2Z_0+Z_0^TB_2^T&B_2Z_{11} &\hdots &B_2Z_{1K}\\
  *^T&*^T&*^T&0&\hdots &0\\
  \vdots &\vdots&\vdots&\vdots&\ddots &\vdots\\
  *^T&*^T&*^T&*^T&\hdots &0}\\
&  = \bmat{
-\frac{\gamma}{\tau_K} I &\frac{1}{\tau_K}D_1&\frac{1}{\tau_K}C_0P+\sum_i C_i Q_i(-\tau_i)^T+\frac{1}{\tau_K}D_2Z_0 &C_1 S_1(-\tau_1)+\frac{1}{\tau_K}D_2Z_{11} &\hdots &C_K S_K(-\tau_K)+\frac{1}{\tau_K}D_2Z_{1K}\\
  *^T&-\frac{\gamma}{\tau_K} I &B_1^T&0&\hdots&0\\
  *^T&*^T&B_2Z_0+Z_0^TB_2^T+L_0+L_0^T&\tau_K A_1S_1(-\tau_1)+B_2Z_{11} &\hdots &\tau_K A_KS_K(-\tau_K)+B_2Z_{1K}\\
  *^T&*^T&*^T&-S_1(-\tau_1)&\hdots &0\\
  \vdots &\vdots&\vdots&\vdots&\ddots &\vdots\\
  *^T&*^T&*^T&*^T&\hdots &-S_k(-\tau_K)}\\
\end{align*}
and
\begin{align*}
&E_i(s)=E_{1i}(s)+E_{2i}(s)+E_{3i}(s)\\
&=\bmat{0\\0\\A_0 Q_i(s) +\dot Q_i(s)+\sum_{j=1}^K A_j R_{ji}(-\tau_j,s) \\ 0 \\ \vdots \\0}+\frac{1}{\tau_K}\bmat{C_0Q_i(s)+\sum_j C_j R_{ji}(-\tau_j,s)\\0\\0\\0\\ \vdots \\ 0}+\frac{1}{\tau_K}\bmat{D_2Z_{2i}(s) \\ 0 \\   \tau_K B_2Z_{2i}(s)    \\0\\ \vdots \\ 0}\\
&=
\frac{1}{\tau_K}\bmat{C_0Q_i(s)+\sum_j C_j R_{ji}(-\tau_j,s)+D_2Z_{2i}(s)\\0\\ \tau_K\left(A_0 Q_i(s) +\dot Q_i(s)+\sum_{j=1}^K A_j R_{ji}(-\tau_j,s)+B_2Z_{2i}(s)\right)  \\ 0 \\ \vdots \\0}.
\end{align*}
We conclude, therefore, that for any $\mbf z \in X$,
\begin{align*}
  &\ip{ \mcl A  \mcl P \mbf z}{\mbf z}_Z+\ip{\mbf z}{ \mcl A  \mcl P \mbf z}_Z+\ip{ \mcl B_2  \mcl Z \mbf z}{\mbf z}_Z+\ip{\mbf z}{ \mcl B_2  \mcl Z \mbf z}_Z + \ip{\mbf z}{\mcl B_1 w}_Z + \ip{\mcl B_1 w}{\mbf z}_Z\notag\\
& - \gamma w^Tw +v^T(\mcl{C}\mcl{P}\mbf z)+(\mcl{C} \mcl P \mbf z)^Tv+v^T(\mcl{D}_2\mcl{Z}\mbf z)+(\mcl{D}_2 \mcl Z \mbf z)^Tv + v^T(D_1 w) + (D_1w)^Tv - \gamma v^Tv\\
&=\ip{\bmat{h \\ \mbf z_{2i}}}{ \mathcal{P}_{\{D,E_i,\dot S_i,G_{ij}\}}\bmat{h \\ \mbf z_{2i}}}_{Z_{m_0,n,K}}\\
&\le- \epsilon \norm{\bmat{\mbf z_{1} \\ \mbf z_{2i}}}_{Z_{n,K}}^2=- \epsilon \norm{\mbf z}_{Z_{n,K}}^2.
\end{align*}
Thus, by Lemma~\ref{lem:selfadjoint_MD} and Theorem~\ref{thm:Hinf_DPS}, we have that for any $w \in L_2$, if  $ x(t)$ and $y(t)$ satisfy Eqn.~\eqref{eqn:MDS}, $\norm{y}_{L_2}\le \gamma \norm{w}_{L_2}$.

\end{proof}

Theorem~\ref{thm:synthesis_MD} provides a convex formulation of the controller synthesis problem for systems with multiple delays. However, the theorem does not provide a way to enforce the operator inequalities or reconstruct the optimal controller. In Section~\ref{sec:SOS} we will review how the operator inequalities can be represented using LMIs. In Sections~\ref{sec:inverse} and~\ref{sec:controller_construction}, we discuss how to invert operators of the $\mcl P_{\{P,Q_i,S_i,R_{ij}\}}$ class and reconstruct the controller gains in a numerically reliable manner.

\section{Enforcing Operator Inequalities in the $\mcl P_{\{P,Q_i,S_i,R_{ij}\}}$ Framework}\label{sec:SOS}
The problem of enforcing operator positivity on $Z_{m,n,K}$ in the $\mcl P_{\{P,Q_i,S_i,R_{ij}\}}$ framework was solved in~\cite{peet_2019} by using a two-step approach. First, we construct an operator $\mcl P_{\{\tilde P,\tilde Q, \tilde S, \tilde R\}}$ whose positivity on $Z_{m,nK,1}$ is equivalent to positivity of the original operator on $Z_{m,n,K}$. Then, assuming that $\tilde Q, \tilde R, \tilde S$ are polynomials, we give an LMI condition on $\tilde P$ and the coefficients of $\tilde Q, \tilde R, \tilde S$ which ensures positivity of $\mcl P_{\{\tilde P,\tilde Q, \tilde S, \tilde R\}}$ on $Z_{m,nK,1}$. Because the transformation from $\{P,Q_i,R_{ij},S_i\}$ to $\{\tilde P,\tilde Q, \tilde R, \tilde S\}$ is linear, if $Q_i,R_{ij},S_i$ are polynomials, the result is an LMI constraint of the coefficients of these original polynomials. For ease of implementation, these two results are combined in single Matlab function which is described in Section~\ref{sec:validation}.

First, we give the following transformation. Specifically, we say that
\begin{equation}
\{\tilde P,\tilde Q,\tilde S,\tilde R\}:=\mathcal{L}_1(P,Q_i,S_i,R_{ij})\label{eqn:L1}\vspace{-2mm}
\end{equation}
if  $a_i=\frac{\tau_i}{\tau_K}$,  $\tilde P=P$ and \vspace{-2mm}
\begin{align*}
&\tilde Q(s) := \bmat{\sqrt{a_1} Q_1(a_1s )& \cdots &\sqrt{a_K} Q_K(a_Ks)}\\
&\tilde S(s) := \bmat{ S_1(a_1s)&0&0\\0&\ddots &0\\0&0& S_K(a_K s)}\\
&  \tilde R(s,\theta):=\\
   & \bmat{\sqrt{a_1 a_1}R_{11}\left(s a_1,\theta a_1\right)&\cdots &\hspace{-2mm}\sqrt{a_1 a_K}R_{1K}\left(s a_1,\theta a_K\right)\\
  \vdots & \cdots & \vdots \\
  \sqrt{a_K a_1} R_{K1}\left(s a_K,\theta a_1\right)&\cdots &\hspace{-2mm}\sqrt{a_K a_K}R_{KK}\left(s a_K,\theta a_K\right)}.
\end{align*}

Then we have the following result~\cite{peet_2019}.

\begin{lem}\label{lem:ZtoL2}
Let $\{\tilde P,\tilde Q,\tilde S,\tilde R\}:=\mathcal{L}_1(P,Q_i,S_i,R_{ij})$. Then
\[
\ip{\bmat{x\\ \phi_i}}{\mathcal{P}_{\{P,Q_i,S_i,R_{ij}\}}\bmat{x\\ \phi_i}}_{Z_{m,n,K}}\ge \alpha \norm{\bmat{ x\\ \phi_i}}_{Z_{m,n,K}}\vspace{-2mm}
\]
for all $\bmat{ x\\ \phi_i} \in Z_{m,n,K}$ if and only if\vspace{-2mm}
\[
\ip{\bmat{x \\ \tilde \phi}}{ \mcl P_{\{\tilde P,\tilde Q,\tilde R,\tilde S\}} \bmat{x \\ \tilde \phi}}_{Z_{m,nK,1}}\ge \alpha \norm{\bmat{\tilde x\\ \tilde \phi}}_{Z_{m,nK,1}}\vspace{-2mm}
\]
 for all $\bmat{\tilde x \\ \tilde \phi} \in Z_{m,nK,1}$.\vspace{-2mm}
\end{lem}\vspace{2mm}

To enforce positivity of $\mcl P_{\{\tilde P,\tilde Q, \tilde S, \tilde R\}}$ on $Z_{m,nK,1}$ as an LMI, we use the following result~\cite{peet_2019}.

\begin{thm}\label{thm:pos_op_joint}
For any functions $Y_1: [-\tau_K,0] \rightarrow \R^{m_1 \times n}$ and $Y_2: [-\tau_K,0] \times [-\tau_K,0] \rightarrow \R^{m_2 \times n}$, square integrable on $[-\tau_K,0]$ with $g(s)\ge 0$ for $s \in [-\tau_K,0]$, suppose that
\begin{align*}
  P & =  M_{11}\cdot \frac{1}{\tau_K}\int_{-\tau_K}^0 g(s)ds \\
  Q(s) & = \frac{1}{\tau_K}\left( g(s)M_{12}Y_1(s)+\int_{-\tau_K}^0 g(\eta)M_{13}Y_2(\eta,s)d\eta \right)\\
  S(s) & = \frac{1}{\tau_K} g(s)Y_1(s)^T M_{22}Y_1(s)\\
  R(s,\theta) & =g(s)Y_1(s)^T M_{23} Y_2(s,\theta)+ g(\theta)Y_2(\theta,s)^T M_{32}Y_1(\theta)+\int_{-\tau_K}^0 g(\eta) Y_2(\eta,s)^T  M_{33}Y_2(\eta,\theta) d\eta
\end{align*}
where $M_{11} \in \R^{m \times m}$, $M_{22} \in \R^{m_1 \times m_1}$, $M_{33} \in \R^{m_2 \times m_2}$ and
  \[
  M=\bmat{M_{11}&M_{12}&M_{13}\\M_{21}&M_{22}&M_{23}\\M_{31}&M_{32}&M_{33}}\ge 0.
  \]

Then $\ip{\mbf x}{\mcl P_{\{P,Q,S,R\}} \mbf x}_{Z_{m,n,1}} \ge 0$ for all $\mbf x \in Z_{m,n,1}$.
\end{thm}
For notational convenience, we use $\{P,Q,S,R\}\in \Xi_{d,m,n}$ to denote the LMI constraints associated with Theorem~\ref{thm:pos_op_joint} as

\begin{align*}
&\Xi_{d,m,n}:=\\
&\left\{\{P,Q,R,S\}:  \substack{ \{P,Q,S,R\}=\{P_1,Q_1,S_1,R_1\}+\{P_2,Q_2,S_2,R_2\},\,\\ \text{ where $\{P_1,Q_1,S_1,R_1\}$ and $\{P_2,Q_2,S_2,R_2\}$ satisfy}\\ \text{ Thm.~\ref{thm:pos_op_joint} with $g=1$ and $g=-s(s+\tau_K)$, respectively.}}  \right\}
\end{align*}
We now have the single unified result:
\begin{cor}\label{cor:positivity_summary}
Suppose there exist $d \in \N$, constant $\epsilon>0$, matrix $P\in \R^{m\times m}$, polynomials $Q_i$, $S_i$,  $R_{ij}$ for $i,j \in [K]$ such that
\[
\mathcal{L}_1(P,Q_i,S_i,R_{ij})\in \Xi_{d,m,nK}.
\]
Then $\ip{\mbf x}{\mathcal{P}_{\{P,Q_i,S_i,R_{ij}\}} \mbf x}_{Z_{m,n,K}} \ge 0 $ for all $\mbf x \in Z_{m,n,K}$.
\end{cor}

A more detailed discussion of these LMI-based methods can be found in~\cite{peet_2019}.

%
%
%
%
%
%
%
%
%
%
%


\section{An Analytic Inverse of $\mcl P_{\{P,Q_i,S_i,R_{ij}\}}$}\label{sec:inverse}
Having taken $Q_i,R_{ij},S_i$ to be polynomials and having given an LMI which enforces strict positivity of the operator $\mcl P_{\{P,Q_i,S_i,R_{ij}\}}$, we now give an analytical representation of the inverse of operators of this class.  The inverse of $\mcl P_{\{P,Q_i,S_i,R_{ij}\}}$ is also of the form $\mcl P_{\{\hat P,\hat Q_i,\hat S_i, \hat R_{ij}\}}$ where expressions for the matrix $\hat P$ and functions $\hat Q_i, \hat R_{ij},\hat S_i$ are given in the following theorem, which is a generalization of the result in~\cite{miao_2017IFAC} to the case of multiple delays. In this result, we first extract the coefficients of the polynomials $Q_i$ and $R_{ij}$ as $Q_i(s) = H_iZ(s)$ and $R_{ij}(s,\theta)=Z(s)^T \Gamma_{ij}Z(\theta)$ where $Z(s)$ is a vector of bases for vector-valued polynomials (typically a monomial basis). The theorem then gives an expression for the coefficients of $\hat Q_i$ and $\hat R_{ij}$ using a similar representation. Note that the results of the theorem are still valid even if the basis functions in $Z(s)$ are not monomials or even polynomials.

\begin{thm}\label{thm:inverse}
  Suppose that $Q_i(s) = H_iZ(s)$ and $R_{ij}(s,\theta)=Z(s)^T \Gamma_{ij}Z(\theta)$ and $\mcl P:=\mathcal{P}_{\{P,Q_i,S_i,R_{ij}\}}$ is a coercive operator where $\mcl P: X \rightarrow X$ and $\mcl P=\mcl P^*$. Define
 \begin{align*}
  H&=\bmat{H_1& \hdots & H_K} \qquad \text{and}\qquad  \Gamma=\bmat{ \Gamma_{11} & \hdots &  \Gamma_{1K}\\ \vdots &&\vdots \\  \Gamma_{K,1}& \hdots &  \Gamma_{K,K}}.
\end{align*}
  Now let
\begin{align*}
K_i&=\int_{-\tau_i}^0  Z(s)S_i(s)^{-1} Z(s)^Tds\\
K&=\bmat{K_1 &0&0\\0&\ddots &0\\0&0&K_K}\\
 \hat H   &=P^{-1} H \left( K H^T P^{-1} H -I- K\Gamma\right)^{-1}\\
 \hat \Gamma&=-(\hat H^T H+\Gamma)(I + K \Gamma)^{-1}\\
  \bmat{\hat H_1& \hdots & \hat H_K}&=\hat H,\qquad \bmat{\hat \Gamma_{11} & \hdots & \hat \Gamma_{1K}\\ \vdots &&\vdots \\ \hat \Gamma_{K,1}& \hdots & \hat \Gamma_{K,K}}=\hat \Gamma.
\end{align*}
If we define
  \begin{align*}
    \hat P & = \left(I- \hat H K H^T\right)P^{-1} \\
    \hat Q_i(s) & =\hat H_i Z(s)S_i(s)^{-1} \\
    \hat S_i(s) & = S_i(s)^{-1} \\
    \hat R_{ij}(s,\theta)&=S_i(s)^{-1} Z(s)^T\hat \Gamma_{ij}Z(\theta)S_j(\theta)^{-1},
  \end{align*}
  then for $\hat{\mcl P}:=\mathcal{P}_{\left\{\hat P,\frac{1}{\tau_K}\hat Q_i,\frac{1}{\tau_K^2}\hat S_i,\frac{1}{\tau_K}\hat R_{ij}\right\}}$, we have that $\hat {\mcl P}=\hat {\mcl P}^*$, $\hat {\mcl P}:X\rightarrow X$, and $\hat {\mcl P}\mcl P \mbf x=\mcl P\hat {\mcl P} \mbf x=\mbf x$ for any $\mbf x  \in Z_{m,n,K}$.
\end{thm}

\begin{proof}
One approach to proving this theorem is to let $\hat{\mcl P}$ be as defined and show that this implies $\hat {\mcl P}\mcl P \mbf x= \mbf x$ for any $x \in Z_{m,n,K}$. Although this is clearly the most direct path towards establishing the theorem statement, it is not the easiest to understand, due to the intensely algebraic nature of the calculations. Thus, in order to allow the reader to understand the derivation of the results and encourage generalization, we will, as much as possible, show how these results were obtained. Specifically, we start by assuming the inverse has the following structure.
\begin{align*}
&\left(\hat{\mathcal{P}} \bmat{x\\ \phi_i}\right)(s) := \\
&\bmat{  \hat P x + \frac{1}{\tau_K} \sum_{i=1}^K \int_{-\tau_i}^0 \hat Q_i(s)\phi_i(s) d s \\
 \hat Q_i(s)^T x +\hspace{-.5mm} \frac{1}{\tau_K} \hat S_i(s)\phi_i(s) + \frac{1}{\tau_K} \sum\limits_{j=1}^K \int_{-\tau_j}^0  \hat R_{ij}(s,\theta)\phi_j(\theta)\, d \theta. }
\end{align*}
Our approach to finding $\hat P, \hat Q_i, \hat S_i$ and $\hat R_{ij}$ is then to calculate $\mbf y=\hat{\mcl P} \mcl P\mbf x$ and use the 5 equality constraints implied by $\mbf y=\mbf x$ to solve for the variables $\hat P, \hat Q_i, \hat S_i$ and $\hat R_{ij}$. To do this, we define
\[
\mbf y(s):=\bmat{y \\ \psi_i(s)}:=\left(\hat{\mathcal{P}}\mathcal{P} \bmat{x\\ \phi_i}\right)(s)
\]
and start by expanding the first term $y=\mbf y_1$.{
\begin{align*}
%
&y = \hat P P x +  \sum_{i=1}^K \int_{-\tau_i}^0 \hat P Q_i(s)\phi_i(s) d s \\
&+   \sum_{i=1}^K \int_{-\tau_i}^0 \hat Q_i(s) Q_i(s)^T xds +  \sum_{i=1}^K \int_{-\tau_i}^0 \hat Q_i(s) S_i(s)\phi_i(s)ds +  \sum_{i=1}^K \int_{-\tau_i}^0 \hat Q_i(s)\sum_{j=1}^K \int_{-\tau_j}^0  R_{ij}(s,\theta)\phi_j(\theta)\, d \theta d s\\
& = \left(\hat P P + \sum_{i=1}^K \int_{-\tau_i}^0 \hat Q_i(s) Q_i(s)^T ds \right) x +  \sum_{i=1}^K \int_{-\tau_i}^0 \left( \hat P Q_i(s)+\hat Q_i(s) S_i(s)\right)\phi_i(s) d s +  \sum_{j=i}^K \int_{-\tau_i}^0 \sum_{j=1}^K \int_{-\tau_j}^0 \hat Q_j(\theta) R_{ji}(\theta,s)\phi_i(s)\,  d s d \theta\\
& = \left(\hat P P + \sum_{i=1}^K \int_{-\tau_i}^0 \hat Q_i(s) Q_i(s)^T ds \right) x +  \sum_{i=1}^K \int_{-\tau_i}^0 \bbbl( \hat P Q_i(s)+\hat Q_i(s) S_i(s) +\sum_{j=1}^K \int_{-\tau_j}^0 \hat Q_j(\theta) R_{ji}(\theta,s)d \theta\bbbr)\phi_i(s) d s
\end{align*}}

From this expansion, we conclude that a sufficient condition for $y=x$ (i.e. $\mbf y_1=\mbf x_1$) is that
\begin{align*}
\hat P P + \sum_{i=1}^K \int_{-\tau_i}^0 \hat Q_i(s) Q_i(s)^T ds &=I, \\
\hat P Q_i(s)+\hat Q_i(s) S_i(s)+\sum_{j=1}^K \int_{-\tau_j}^0 \hat Q_j(\theta) R_{ji}(\theta,s)d \theta&=0
\end{align*}
for all $i \in [K]$. This provides two sets of equality constraints which will help us determine $\hat P$ and $\hat Q$.
We next examine the more complicated terms $\psi_i=\mbf y_2$.{\small
\begin{align*}
&\psi_i(s)=\hat Q_i(s)^T P x +  \sum_{j=1}^K \int_{-\tau_j}^0 \hat Q_i(s)^T Q_j(\theta)\phi_j(\theta) d \theta+ \hat S_i(s) Q_i(s)^T x + \hat S_i(s) S_i(s)\phi_i(s) +  \sum_{j=1}^K \int\limits_{-\tau_j}^0 \hspace{-2mm} \hat S_i(s) R_{ij}(s,\theta)\phi_j(\theta)\, d \theta \\
&+ \left( \sum_{j=1}^K \int_{-\tau_j}^0  \hat R_{ij}(s,\theta) Q_j(\theta)^T d\theta\right) x+ \sum_{j=1}^K \int_{-\tau_j}^0  \hat R_{ij}(s,\theta) S_j(\theta)\phi_j(\theta)d \theta +  \sum_{j=1}^K \int_{-\tau_j}^0  \hat R_{ij}(s,\theta) \sum_{k=1}^K \int_{-\tau_k}^0  R_{jk}(\theta,\eta)\phi_k(\eta)\, d \eta d \theta\\
&=\left(\hat Q_i(s)^T P + \hat S_i(s) Q_i(s)^T + \sum_{j=1}^K \int_{-\tau_j}^0  \hat R_{ij}(s,\theta) Q_j(\theta)^T d\theta\right) x  +  \hat S_i(s) S_i(s)\phi_i(s)\\
& +  \sum_{j=1}^K \int_{-\tau_j}^0   \bbbl(\hat Q_i(s)^T Q_j(\theta)+\hat S_i(s) R_{ij}(s,\theta)+\hat R_{ij}(s,\theta) S_j(\theta)\bbbr)\phi_j(\theta)\, d \theta  + \sum_{j=1}^K \int_{-\tau_j}^0 \sum_{k=1}^K \int_{-\tau_k}^0  \hat R_{ik}(s,\eta)   R_{kj}(\eta,\theta) \, d \eta \phi_j(\theta) d \theta\\
&=\left(\hat Q_i(s)^T P + \hat S_i(s) Q_i(s)^T + \sum_{j=1}^K \int_{-\tau_j}^0  \hat R_{ij}(s,\theta) Q_j(\theta)^T d\theta\right) x  +  \hat S_i(s) S_i(s)\phi_i(s)\\
& +  \sum_{j=1}^K \int_{-\tau_j}^0  \bbbl(\hat Q_i(s)^T Q_j(\theta)+\hat S_i(s) R_{ij}(s,\theta)+\hat R_{ij}(s,\theta) S_j(\theta)+ \sum_{k=1}^K \int_{-\tau_k}^0  \hat R_{ik}(s,\eta)   R_{kj}(\eta,\theta) \, d \eta\bbbr)\phi_j(\theta)\, d \theta
\end{align*}}

From this expansion, we conclude that a sufficient condition for $\psi_i(s)=\phi_i(s)$ (i.e. $\mbf y_2=\mbf x_2$) is that
\begin{align*}
& \hat S_i(s) S_i(s)=I\\
&\hat Q_i(s)^T P + \hat S_i(s) Q_i(s)^T + \sum_{j=1}^K \int_{-\tau_j}^0  \hat R_{ij}(s,\theta) Q_j(\theta)^T d\theta=0\\
&\hat Q_i(s)^T Q_j(\theta)+\hat S_i(s) R_{ij}(s,\theta)+\hat R_{ij}(s,\theta) S_j(\theta)\\
&\qquad \qquad \qquad  + \sum_{k=1}^K \int_{-\tau_k}^0  \hat R_{ik}(s,\eta)   R_{kj}(\eta,\theta) \, d \eta=0.
\end{align*}


We now have 5 constraints which $\hat P, \hat Q_i, \hat S_i$ and $\hat R_{ij}$ must satisfy if $\hat{\mcl P}$ is to be an inverse of $\mcl P$:
\begin{align*}
& \hat S_i(s) S_i(s)=I\\
&\hat P P + \sum_{i=1}^K \int_{-\tau_i}^0 \hat Q_i(s) Q_i(s)^T ds =I\\
&\hat P Q_i(s)+\hat Q_i(s) S_i(s)+\sum_{j=1}^K \int_{-\tau_j}^0 \hat Q_j(\theta) R_{ji}(\theta,s)d \theta=0 \qquad \forall i\\
&\hat Q_i(s)^T P + \hat S_i(s) Q_i(s)^T + \sum_{j=1}^K \int_{-\tau_j}^0  \hat R_{ij}(s,\theta) Q_j(\theta)^T d\theta=0\\
&\hat Q_i(s)^T Q_j(\theta)+\hat S_i(s) R_{ij}(s,\theta)+\hat R_{ij}(s,\theta) S_j(\theta)\\
&\qquad \qquad \qquad  + \sum_{k=1}^K \int_{-\tau_k}^0  \hat R_{ik}(s,\eta)   R_{kj}(\eta,\theta) \, d \eta=0
\end{align*}
If all 5 constraints are satisfied, we can conclude that $\hat{\mcl P}\mcl P\mbf x=\mbf x$. Clearly, the first constraint is satisfied if $\hat S_i(s)=S_i(s)^{-1}$.

We now parameterize the variables $\hat Q_i$ and $\hat R_{ij}$ as
\begin{align*}
  \hat Q_i(s) & =\hat H_iZ(s)\hat S_i(s),\,\;  \hat R_{ij}(s,\theta)=\hat S_i(s)^T Z(s)\hat \Gamma_{ij}Z(\theta)\hat S_j(\theta)\\
\end{align*}
and examine the second constraint, which is equivalent to
\[
\hat P P  =I- \sum_{i=1}^K \int_{-\tau_i}^0 \hat Q_i(s) Z(s)^Tds H_i^T.
\]
Solving this expression for $\hat P$ in terms of $\hat H$, we obtain
\begin{align*}
\hat P &=\left(I- \sum_{i=1}^K \int_{-\tau_i}^0 \hat Q_i(s) Z(s)^Tds H_i^T\right)P^{-1}\\
&=\left(I- \sum_{i=1}^K \hat H_i\left(\int_{-\tau_i}^0 Z(s)\hat S_i(s) Z(s)^Tds\right) H_i^T\right)P^{-1}\\
&=\left(I- \sum_{i=1}^K \hat H_iK_i H_i^T\right)P^{-1}=\left(I- \hat H K H^T\right)P^{-1}.\\
\end{align*}
We now examine the third set of constraints, indexed by $i\in [K]$:{\small
\begin{align*}
&\hat P Q_i(s)+\hat Q_i(s) S_i(s)+\sum_{j=1}^K \int_{-\tau_j}^0 \hat Q_j(\theta) R_{ji}(\theta,s)d \theta\\
&=\hat P H_i Z(s)+ \hat H_i Z(s)+\sum_{j=1}^K \hat H_j \int_{-\tau_j}^0  Z(\theta)\hat S_j(\theta) Z(\theta)^T \Gamma_{ji}Z(s)d \theta\\
&=\left(\hat P H_i + \hat H_i +\sum_{j=1}^K \hat H_j K_j \Gamma_{ji}\right)Z(s)=0.
\end{align*}}
Combining these $K$ constraints into a single expression yields
\[
\hat P H + \hat H +\hat H K \Gamma=0.
\]
Substituting our expression for $\hat P$ now yields the constraint

\begin{align*}
&\hat P H + \hat H +\hat H K \Gamma\\
&=\left(I- \hat H K H^T\right)P^{-1} H + \hat H +\hat H K \Gamma\\
&=P^{-1} H- \hat H \left( K H^T P^{-1} H -I- K\Gamma\right)  =0,
\end{align*}
which is equivalent to
\[
 \hat H   =P^{-1} H \left( K H^T P^{-1} H -I- K\Gamma\right)^{-1}.
\]
Thus we have found an expression for $\hat H$. Furthermore, since we have already found an expression for $\hat{\mcl P}$ in terms of $\hat H$, all that now remains is to solve for $\hat \Gamma$. For this result, we turn to the 5th set of constraints:
\begin{align*}
&\hat Q_i(s)^T Q_j(\theta)+\hat S_i(s) R_{ij}(s,\theta)+\hat R_{ij}(s,\theta) S_j(\theta) + \sum_{k=1}^K \int_{-\tau_k}^0  \hat R_{ik}(s,\eta)   R_{kj}(\eta,\theta) \, d \eta\\
&=\hat S_i(s)^T Z(s)^T \hat H_i^T H_j Z(\theta)+\hat S_i(s) Z(s)^T \Gamma_{ij}Z(\theta)+\hat S_i(s)^T Z(s)\hat \Gamma_{ij}Z(\theta)\\
& + \sum_{k=1}^K \int_{-\tau_k}^0 \hat S_i(s)^T Z(s)^T\hat \Gamma_{ik}Z(\eta)\hat S_k(\eta)  Z(\eta) d \eta \Gamma_{kj}Z(\theta) \\
&=\hat S_i(s)^T Z(s)^T \left( \hat H_i^T H_j + \Gamma_{ij}+\hat \Gamma_{ij} + \sum_{k=1}^K  \hat \Gamma_{ik} K_k \Gamma_{kj}\right)Z(\theta)\\
& =0 \qquad \forall i,j \in [K]
\end{align*}
Combining these $K^2$ constraints into a single expression yields
\[
\hat H^T H + \Gamma+\hat \Gamma + \hat \Gamma K \Gamma=0.
\]
Solving this expression for $\hat \Gamma$, we find
\[
\hat \Gamma=-(\hat H^T H+\Gamma)(I + K \Gamma)^{-1}.
\]
We have now derived expressions for $\hat P$, $\hat S$, $\hat H$, and $\hat \Gamma$. However, to show that $\hat{\mcl P}\mcl P\mbf x=\mbf x$, we must verify that the fourth constraint is also satisfied. Namely,
\begin{align*}
&\hat Q_i(s)^T P + \hat S_i(s) Q_i(s)^T + \sum_{j=1}^K \int_{-\tau_j}^0  \hat R_{ij}(s,\theta) Q_j(\theta)^T d\theta\\
&=\hat S_i(s)  Z(s)^T \hat H_i^T P + \hat S_i(s)  Z(s)^T H_i^T + \sum_{j=1}^K \int_{-\tau_j}^0  \hat S_i(s)  Z(s)^T \hat \Gamma_{ij}  Z(\theta) \hat S_j(\theta) Z(\theta)^T d\theta H_j^T\\
&=\hat S_i(s)  Z(s)^T \left( \hat H_i^T P +  H_i^T + \sum_{j=1}^K    \hat \Gamma_{ij}  K_j H_j^T\right)=0\qquad
\end{align*}
for all $i\in [K]$, which is satisfied if
\[
\hat H_i^T P +  H_i^T + \sum_{j=1}^K    \hat \Gamma_{ij}  K_j H_j^T=0 \qquad \forall i\in [K].
\]
Combining these $K$ constraints into a single expression yields
\[
 \hat H^T P +  H^T +   \hat \Gamma  K H^T=0.
\]
To verify this is satisfied, we let $L=H^TP^{-1}H$ and $T=\left( I+K\Gamma-K L \right)^{-1}$. Then $\hat H=-P^{-1}HT$ and thus
\begin{align*}
& \hat H^T P +  H^T +   \hat \Gamma  K H^T\\
&=-T^T H^T P^{-1}P  +  H^T +   \hat \Gamma  K H^T\\
&=\left( -T^T   +  I +   \hat \Gamma  K \right)H^T.
\end{align*}
Substituting in $\hat \Gamma=(T^TL-\Gamma)(I+K\Gamma)^{-1}$, and observing that $\Gamma$, $L$ and $K$ are symmetric (for $\Gamma$, this is due to $\mcl P=\mcl P^{*}$), we have that
\begin{align*}
&-T^T   +  I +   \hat \Gamma  K \\
&=I-T^T +   (T^TL-\Gamma)(I+K\Gamma)^{-1}  K\\
&=I - T^T(I+\Gamma K)(I+\Gamma K)^{-1}   +   (T^TL-\Gamma)K(I+\Gamma K)^{-1}  \\
&=I - T^T(I+\Gamma K)(I+\Gamma K)^{-1}   +   T^TL K(I+\Gamma K)^{-1}\\
&\qquad\qquad\qquad\qquad\qquad\qquad\qquad\qquad -\Gamma K(I+\Gamma K)^{-1}  \\
&=I - T^T(I+\Gamma K+LK)(I+\Gamma K)^{-1}   -\Gamma K(I+\Gamma K)^{-1}  \\
&=I - (I+\Gamma K)^{-1}   -\Gamma K(I+\Gamma K)^{-1}  \\
&=I - (I+\Gamma K)(I+\Gamma K)^{-1}    =I - I=0    \\
\end{align*}
In a similar manner, it can be shown that $\mcl P \hat{\mcl P}\mbf x=\mbf x$. It can be likewise shown directly that $\hat{\mcl P}:X\rightarrow X$ through a lengthy series of algebraic manipulations. However, this property is also established by Theorem~\ref{thm:dual}.
\end{proof}

\section{Controller Reconstruction and Numerical Implementation} \label{sec:controller_construction}
In this section, we reconstruct the controller using $\mcl Z$ and $\mcl P^{-1}$ and explain how this can be implemented numerically. First, we have the following obvious result.
\begin{lem} \label{lem:composition}
Suppose that
\begin{align}
&\left(\mcl Z \bmat{y\\ \psi_i}\right):=\bmat{Z_0 y + \sum_i Z_{1i} \psi_i(-\tau_i)+ \sum_i\int_{-\tau_i}^0 Z_{2i}(s) \psi_i(s) ds}\label{eqn:Z_defn}
\end{align}
and
\begin{align*}
&\left(\hat{\mathcal{P}} \bmat{x\\ \phi_i}\right)(s) := \bmat{  \hat P x + \frac{1}{\tau_K} \sum_{i=1}^K \int_{-\tau_i}^0 \hat Q_i(s)\phi_i(s) d s \\
 \hat Q_i(s)^T x +\hspace{-.5mm} \frac{1}{\tau_K} \hat S_i(s)\phi_i(s) + \frac{1}{\tau_K} \sum\limits_{j=1}^K \int_{-\tau_j}^0  \hat R_{ij}(s,\theta)\phi_j(\theta)\, d \theta. }
\end{align*}
Then if $u(t)=\mcl Z \hat P \mbf x(t)$,
\[
u(t)=K_0 x(t) + \sum_i K_{1i} x(t-\tau_i)+ \sum_i\int_{-\tau_i}^0 K_{2i}(s) x(t+s) ds
\]
where
\begin{align*}
  &K_0  = Z_0 \hat P +\sum_j \left( Z_{1j} \hat Q_j(-\tau_j)^T+\int_{-\tau_j}^0 Z_{2j}(s)\hat Q_j(s)^T ds\right)\\
  &K_{1i}  =\frac{1}{\tau_K}  Z_{1i} \hat S_i(-\tau_i) \\
  &K_{2i}(s)  =\frac{1}{\tau_K}\bbbl( Z_0 \hat Q_i(s)+Z_{2i}(s)\hat S_i(s) +\sum_{j=1}^K \bbl(Z_{1j} \hat R_{ji}(-\tau_j,s)+ \int_{\theta=-\tau_j}^0 Z_{2j}(\theta)\hat R_{ji}(\theta,s)d \theta\bbr)\bbbr)
\end{align*}
\end{lem}
\begin{proof}
Suppose that $\mcl K=\mcl Z \mcl P^{-1}=\mcl Z \hat{\mcl P}$ where

{\small
\begin{align*}
&\mcl Z \hat{\mcl P} \bmat{x\\ \phi_i}=Z_0 \left( \hat P x + \frac{1}{\tau_K} \sum_{i=1}^K \int_{-\tau_i}^0 \hat Q_i(s)\phi_i(s) d s\right) +\sum_i Z_{1i} \bbbl(\hat Q_i(-\tau_i)^T x +\hspace{-.5mm} \frac{1}{\tau_K} \hat S_i(-\tau_i)\phi_i(-\tau_i) + \frac{1}{\tau_K} \sum_{j=1}^K \int_{-\tau_j}^0  \hat R_{ij}(-\tau_i,\theta)\phi_j(\theta)\, d \theta\bbbr)\\
&\qquad +\sum_i\int_{-\tau_i}^0 Z_{2i}(s) \bbbl(\hat Q_i(s)^T x +\hspace{-.5mm} \frac{1}{\tau_K} \hat S_i(s)\phi_i(s)  + \frac{1}{\tau_K} \sum_{j=1}^K \int_{\theta=-\tau_j}^0  \hat R_{ij}(s,\theta)\phi_j(\theta)\, d \theta\bbbr) ds\\
&=  Z_0 \hat P x + \frac{1}{\tau_K} \sum_{j=1}^K \int_{-\tau_j}^0 Z_0 \hat Q_j(s)\phi_j(s) d s+ \sum_i Z_{1i} \hat Q_i(-\tau_i)^T x +\hspace{-.5mm} \frac{1}{\tau_K} \sum_i Z_{1i} \hat S_i(-\tau_i)\phi_i(-\tau_i) + \frac{1}{\tau_K} \sum_{j=1}^K \int_{-\tau_j}^0 \sum_{i=1}^K Z_{1i} \hat R_{ij}(-\tau_i,s)\phi_j(s)\, d s\\
&\quad +\sum_i\int\limits_{-\tau_i}^0 Z_{2i}(s)\hat Q_i(s)^T x ds +\frac{1}{\tau_K} \sum_j\int\limits_{-\tau_j}^0 Z_{2j}(s)\hat S_j(s)\phi_j(s)ds+ \frac{1}{\tau_K} \sum_{i,j=1}^K \int_{\theta=-\tau_i}^0 \int_{s=-\tau_j}^0  Z_{2i}(\theta)\hat R_{ij}(\theta,s)\phi_j(s)\, d \theta ds\\
&=  \left(Z_0 \hat P +\sum_i \left( Z_{1i} \hat Q_i(-\tau_i)^T+\int_{-\tau_i}^0 Z_{2i}(s)\hat Q_i(s)^T ds\right) \right)x+ \hspace{-.5mm} \frac{1}{\tau_K} \sum_i Z_{1i} \hat S_i(-\tau_i)\phi_i(-\tau_i)\\
&+\frac{1}{\tau_K} \sum_j\int_{-\tau_j}^0\bbbl( Z_0 \hat Q_j(s)+Z_{2j}(s)\hat S_j(s)+\sum_{i=1}^K \bbl(Z_{1i} \hat R_{ij}(-\tau_i,s)+ \int_{\theta=-\tau_i}^0 Z_{2i}(\theta)\hat R_{ij}(\theta,s)d \theta\bbr)\bbbr)\phi_j(s)ds\\
&=  \left(Z_0 \hat P +\sum_j \left( Z_{1j} \hat Q_j(-\tau_j)^T+\int_{-\tau_j}^0 Z_{2j}(s)\hat Q_j(s)^T ds\right) \right)x + \hspace{-.5mm} \frac{1}{\tau_K} \sum_i Z_{1i} \hat S_i(-\tau_i)\phi_i(-\tau_i)\\
&+\frac{1}{\tau_K} \sum_i\int_{-\tau_i}^0\bbbl( Z_0 \hat Q_i(s)+Z_{2i}(s)\hat S_i(s) +\sum_{j=1}^K \bbl(Z_{1j} \hat R_{ji}(-\tau_j,s)+ \int_{\theta=-\tau_j}^0 Z_{2j}(\theta)\hat R_{ji}(\theta,s)d \theta\bbr)\bbbr)\phi_i(s)ds
\end{align*}}

We conclude that the controller $\mcl K$ has the form
\[
\left(\mcl K \bmat{x\\ \phi_i}\right):=\bmat{K_0 x + \sum\limits_i K_{1i} \phi_i(-\tau_i)+ \sum\limits_i\hspace{-1mm}\int\limits_{-\tau_i}^0 \hspace{-1.5mm}K_{2i}(s) \phi_i(s) ds}
\]
\end{proof}
We conclude that given $\hat P, \hat Q_i, \hat S_i$ and $\hat R_{ij}$, it should be possible to compute the controller gains $K_0$, $K_{1i}$ and $K_{2i}$. In practice, however, if $S$ is polynomial, then $\hat S_i(s) = S(s)^{-1}$ will be a rational matrix-valued function. This implies that $\hat Q_i$ and $\hat R_{ij}$ are likewise rational. Computing and analytically integrating such rational functions poses serious challenges. Fortunately, however, this task can be largely avoided. Specifically, if we use the formulae from Theorem~\ref{thm:inverse} and substitute into the expression for $u(t)$, we obtain the following

\begin{cor}\label{cor:inverse_simple}
  If $\mcl Z$ is as defined in Lemma~\ref{lem:composition} and $\hat{\mcl P}$ is as defined in Theorem~\ref{thm:inverse} and $u(t)=\mcl Z \hat P  \bmat{x(t) \\ x(t+s)}$, then
\[
u(t)=K_0 x(t) + \sum_i K_{1i} x(t-\tau_i)+ \sum_i\int_{-\tau_i}^0 K_{2i}(s) x(t+s) ds
\]
where
\begin{align*}
  &K_0  = Z_0 \hat P +\sum_j \left( Z_{1j} S_j(-\tau_j)^{-1}Z(-\tau_j)^T+O_j\right)\hat H_j^T\\
  &K_{1i}  =\frac{1}{\tau_K}  Z_{1i} S_i(-\tau_i)^{-1} \\
  &K_{2i}(s)  =\frac{1}{\tau_K}\bbbl( \left(Z_0 \hat H_i Z(s)
  +Z_{2i}(s)\right) +\sum_{j=1}^K \left(Z_{1j} S_j(-\tau_j)^{-1} Z(-\tau_j)^T + O_j\right)\hat \Gamma_{ji} Z(s)\bbbr) S_i(s)^{-1}\\
  &O_i=\int_{-\tau_j}^0 Z_{2j}(s) S_j(s)^{-1}Z(s)^T ds
\end{align*}
\end{cor}
The proof is straightforward.

The advantage of this representation is that the matrices $O_i$ can be numerically calculated a priori to machine precision using trapezoidal integration without an analytic expression for $S^{-1}$. Naturally, implementation still requires integration of $\int_{-\tau_i}^0 K_{2i}(s) \phi_i(s) ds$ in real-time. However, practical implementation of such controllers is typically based on sampling $\{t_i\}$ of the history, meaning computation of $\int_{-\tau_i}^0 K_{2i}(s) \phi_i(s) ds$ can be reduced to matrix multiplication based on numerical evaluations of $S(t_i)^{-1}$. This real-time implementation can be further simplified if the state-feedback controller is combined with an $H_\infty$-optimal estimator, as described in~\cite{peet_2018MTNS}.

\section{An LMI Formulation of the $H_\infty$-Optimal Controller Synthesis Problem for Multi-Delay Systems}\label{sec:synthesis_LMI}
In this section, we combine all previous results to give a concise formulation of the controller synthesis problem in the LMI framework.
\begin{thm}\label{thm:synthesisLMI_MD}
For any $\gamma >0$, suppose there exist $d \in \N$, constant $\epsilon>0$, matrix $P\in \R^{n\times n}$, polynomials $S_i, Q_i \in W_2^{n\times n}[T_i^0]$, $R_{ij}\in W_2^{n\times n}\left[T_i^0 \times T_j^0 \right]$ for $i,j \in [K]$, matrices $Z_0,Z_{1i} \in \R^{p \times n}$ and polynomials $Z_{2i}[T_i^0] \in W_2^{p \times n}$ for $i \in [K]$ such that\vspace{-2mm}
\begin{align*}
\mathcal{L}_1(P- \epsilon I_n,Q_i,S_i-\epsilon I_n,R_{ij}) &\in \Xi_{d,n,nK}\\
-\mathcal{L}_1(D\hspace{-.5mm}+\hspace{-.5mm}\epsilon \hat I ,E_i,\dot S_i +\epsilon I_n ,G_{ij}) & \in \Xi_{d,q+m+n(K+1),nK},\\[-7mm]
\end{align*}
where $D$, $E_i$, $G_{ij}$ are as defined in Theorem~\ref{thm:synthesis_MD}, $\hat I=\diag(0_{q+m},I_{n},0_{nK})$, and $\mcl L_1$ is as defined in Eqn.~\eqref{eqn:L1}.
Furthermore, suppose $P,Q_i,S_i,R_{ij}$ satisfy the conditions of Lemma~\ref{lem:selfadjoint_MD}. Let
\[
u(t)=K_0 x(t) + \sum_i K_{1i} x(t-\tau_i)+ \sum_i\int_{-\tau_i}^0 K_{2i}(s) x(t+s) ds
\]
where $\hat P$, $\hat H_i$, and $\hat \Gamma_{ji}$ for $Z(s)$ are as defined in Theorem~\ref{thm:inverse} and
\begin{align*}
  &K_0  = Z_0 \hat P +\sum_j \left( Z_{1j} S_j(-\tau_j)^{-1}Z(-\tau_j)^T+O_j\right)\hat H_j^T\\
  &K_{1i}  =\frac{1}{\tau_K}  Z_{1i} S_i(-\tau_i)^{-1} \\
  &K_{2i}(s)  =\frac{1}{\tau_K}\bbbl( \left(Z_0 \hat H_i Z(s)
  +Z_{2i}(s)\right) +\sum_{j=1}^K \left(Z_{1j} S_j(-\tau_j)^{-1} Z(-\tau_j)^T + O_j\right)\hat \Gamma_{ji} Z(s)\bbbr) S_i(s)^{-1}\\
  &O_i=\int_{-\tau_j}^0 Z_{2j}(s) S_j(s)^{-1}Z(s)^T ds
\end{align*}

Then for any $w \in L_2$, if $y(t)$ and $x(t)$ satisfy Equation~\eqref{eqn:MDS}, $\norm{y}_{L_2}\le \gamma \norm{w}_{L_2}$.

\end{thm}\vspace{2mm}
\begin{proof} Define $\mcl P:=\mathcal{P}_{\{P,Q_i,S_i,R_{ij}\}}$. By assumption, $\mcl P$ satisfies the conditions of Lemma~\ref{lem:selfadjoint_MD}. By Corollary~\ref{cor:positivity_summary}, we have
\begin{align*}
&\ip{\mbf x}{\mathcal{P}_{\{P-\epsilon I_n,Q_i,S_i-\epsilon I_n,R_{ij}\}}\mbf x}_{Z_{n,K}}\\
&=\ip{\mbf x}{\mathcal{P}_{\{P,Q_i,S_i,R_{ij}\}}\mbf x}_{Z_{n,K}} -\epsilon \norm{\mbf x}^2_{Z_{n,K}}\ge 0
\end{align*}
for all $\mbf x \in Z_{n,K}$. Similarly, we have
\begin{align*}
&\ip{\bmat{\bmat{v\\ w \\ \mbf z_1 \\ f} \\ \mbf z_{2i}}}{\mathcal{P}_{\{D+\epsilon \hat I ,E_i,\dot S_i +\epsilon I_n ,G_{ij}\}}\bmat{\bmat{v\\ w \\ \mbf z_1 \\ f} \\ \mbf z_{2i}}}_{\hspace{-3mm}Z_{q+m+n(K+1),n,K}}\\
&=\ip{\bmat{\bmat{v\\ w \\ \mbf z_1 \\ f} \\ \mbf z_{2i}}}{\mathcal{P}_{\{D ,E_i,\dot S_i  ,G_{ij}\}}\bmat{\bmat{v\\ w \\ \mbf z_1 \\ f} \\ \mbf z_{2i}}}_{Z_{q+m+n(K+1),n,K}} \hspace{-20mm}+\epsilon \norm{\bmat{\mbf z_1\\ \mbf z_{2i}}}^2_{Z_{n,K}}\\
&\le 0.\\[-12mm]
\end{align*}
for all $\mbf z_1 \in \R^n$ and $\bmat{\bmat{v\\ w \\ \mbf z_1 \\ f} \\ \mbf z_{2i}} \in Z_{q+m+n(K+1),n,K}$.

Furthermore, by Theorem~\ref{thm:inverse} and Corollary~\ref{cor:inverse_simple}, $u(t)=\mcl Z \mcl P^{-1} \bmat{x(t)\\x(t+s)}$ where{\small
\[
\left(\mcl Z \bmat{x \\ \phi_i}\right)(s):=Z_0 x + \sum_{i=1}^K Z_{1i} \phi_i(-\tau_i) + \sum_{i=1}^K \int_{-\tau_i}^0 Z_{2i}(s)\phi_i(s)ds.\vspace{-2mm}
\]}
Therefore, by Theorem~\ref{thm:synthesis_MD}, if $y(t)$ and $x(t)$ satisfy Equation~\eqref{eqn:MDS}, $\norm{y}_{L_2}\le \gamma \norm{w}_{L_2}$
\end{proof}

\section{Numerical Testing, Validation and Practical Implementation}\label{sec:validation}

The algorithms described in this paper have been implemented in Matlab within the DelayTOOLs framework, which is based on SOSTOOLS and the pvar framework. Several supporting functions were described in~\cite{peet_2019} and these are sufficient to enforce the conditions of Theorem~\ref{thm:synthesisLMI_MD}. For all examples, the computation time is in CPU seconds on an Intel i7-5960X 3.0GHz processor. This time corresponds to the interior-point (IPM) iteration in SeDuMi and does not account for preprocessing, postprocessing, or for the time spent on polynomial manipulations formulating the SDP using SOSTOOLS. Such polynomial manipulations can significantly exceed SDP computation time for small problems.

For simulation and practical use, some additional functionality has been added to facilitate calculation of controller gains and real-time implementation. The most significant new function introduced in this paper is \verb+P_PQRS_Inverse_joint_sep_ndelay+, which takes the matrix $P$ and polynomials $Q_i$, $S_i$, and $R_{ij}$ and computes $\hat P$, $\hat H_i$, and $\hat \Gamma_{ij}$ as described in Theorem~\ref{thm:inverse}. In addition, the script \verb+solver_ndelay_opt_control+ combines all aspects of this paper and simulates the resulting controller in closed loop. For simulation, a fixed-step forward difference method is used, with a different set of states representing each delay channel. In the simulation results given below, 200 spatial discretization points are used for each delay channel.

\subsection{Bounding the $H_\infty$ norm of a Multi-Delay System}
Naturally, the results of this paper can be used to bound the $H_\infty$ norm of a time-delay system by simply setting $B_2=0$. In this subsection, we take this approach and verify that the resulting $H_\infty$ norm bounds are accurate to several decimal places as compared with a high-order Pad\'e-based approximation scheme and compare favorably with existing results in the literature. In each case, the Pad\'e estimate is calculated using a 10th-order Pad\'e approximation combined with the Matlab \verb+norm+ command. The minimum $H_\infty$ norm bound is indicated by $\gamma_{\min}$.

\paragraph{Example A.1}
\begin{align*}
\dot x(t)&=\bmat{-2 &0\\0& -.9}x(t)+\bmat{-1 &0\\-1&-1}x(t-\tau)+\bmat{-.5\\1}w(t)\\
y(t)&=\bmat{1 & 0}x(t)
\end{align*}
\[
\hbox{\begin{tabular}{c|c|c|c|c|c|c}
$d$ & $1$  & $2$ & $3$  & \text{Pad\'e}&\cite{fridman_2001}& \cite{shaked_1998}\\
\hline
$\gamma_{\min}$ & .2373 & .2365 &  .2365  &.2364 &.32 & 2\\
\end{tabular}}
\]

\begin{figure}\vspace{-3mm}
\includegraphics[width=.47\textwidth]{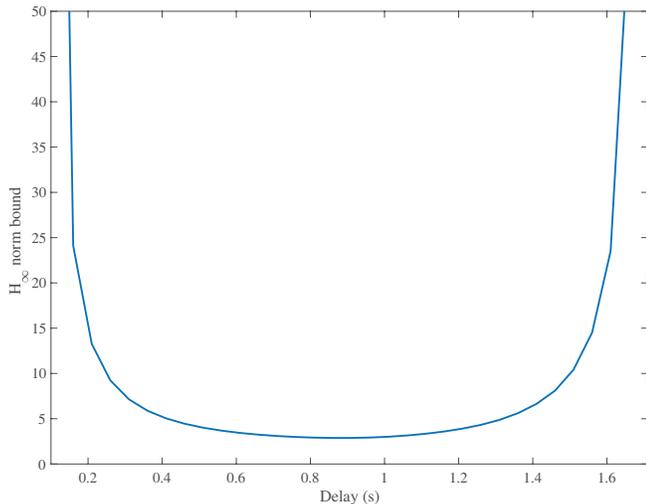}\vspace{-3mm}
\caption{Calculated Open Loop $H_\infty$ norm bound vs. delay for Ex. A.2}\label{fig:ex_a2}
\end{figure}

\paragraph{Example A.2}
In Example A.2, we consider a well-studied example which is known to be stable for delays in the interval $\tau \in [.100173,1.71785]$.
\begin{align*}
\dot x(t)&=\bmat{0 &1\\-2& .1}x(t)+\bmat{0 &0\\1&0}x(t-\tau)+\bmat{1 &0\\0&1}w(t)\\
y(t)&=\bmat{0 & 1}x(t)
\end{align*}
We use the algorithm to compute bounds for the open-loop $H_\infty$ norm of this system as the delay varies within this interval. The results are illustrated in Figure~\ref{fig:ex_a2}. Note that, as expected, the $H_\infty$ norm approaches infinity quickly as we approach the limits of the stable region.

\subsection{Validation of $H_\infty$ optimal controller synthesis}
We now apply the controller synthesis algorithm to several problems. Unfortunately, there are very few challenging example problems available in the literature. When these examples do exist, they are often trivial in the sense that the dynamics can be entirely eliminated by the controller - meaning only the control effort is to be minimized and the achievable norms do not change significantly with delay or other parameters. The problems listed below were found to be the most challenging as measured by either significant variation of the closed-loop norm with delay or the requirement for a degree of more than 1 to achieve optimal performance. In each case, the results are compared to existing results in the literature (when available) and to an $H_\infty$ optimal controller designed for the ODE obtained by using a 10th order Pad\'e approximation of the delay terms.

\paragraph{Example B.1}
\begin{align*}
\dot x(t)&=\bmat{0 &0\\0& 1}x(t)+\bmat{-1 &-1\\0&-.9}x(t-\tau) +\bmat{1\\1}w(t) + \bmat{0\\1}u(t)\\
y(t)&=\bmat{1 & 0\\0 &0}x(t)+\bmat{0\\.1}u(t)
\end{align*}
\[{\small
\hbox{\begin{tabular}{c|c|c|c|c|c|c}
$\gamma_{\min}$ & $d=1$  & $d=2$ & $d=3$  & \text{Pad\'e} & \cite{fridman_2003}&\cite{li_1997}\\
\hline
$\tau=.99$ &  .10001  & .10001 & .10001 &  .1000 & .2284 & 1.882 \\
$\tau=2$ &  1.438  & 1.353 & 1.332 &  1.339 & $\infty$ & $\infty$ \\
\hline
CPU sec & .478 & .879 & 2.48  & 2.78 & N/A &N/A\\
\end{tabular}}}
\]

\paragraph{Example B.2} This example comes from~\cite{fridman_1998}. In that work, the authors set $D_1=D_2=0$ and, for e.g. $\tau=.3$, obtained a closed loop $H_\infty$ bound of $\gamma =.3983$. Theorem~\ref{thm:synthesisLMI_MD} was able to find a closed loop controller for arbitrarily small closed-loop norm bound ($<10^{-6}$). This is because the control effort is not included in the regulated output. We remedy this and add a second regulated output to obtain
\begin{align*}
\dot x(t)&=\bmat{2 &1\\0& -1}x(t)+\bmat{-1 &0\\-1&1}x(t-\tau)+\bmat{-.5\\1}w(t) + \bmat{3\\1}u(t)\\
y(t)&=\bmat{1 & -.5\\0 &0}x(t)+\bmat{0\\1}u(t).
\end{align*}
\[
\hbox{\begin{tabular}{c|c|c|c|c|c}
$\gamma_{\min}$ & $d=1$  & $d=2$ & $d=3$  & \text{Pad\'e} & \cite{fridman_1998}\\
\hline
$\tau=.3$ &  .3953  & .3953 & .3953 &   .3953 & N/A\\
\hline
CPU sec & .655 & 1.248 & 2.72  & N/A & N/A  \\
\end{tabular}}
\]

\begin{figure}\vspace{-3mm}
\includegraphics[width=.47\textwidth]{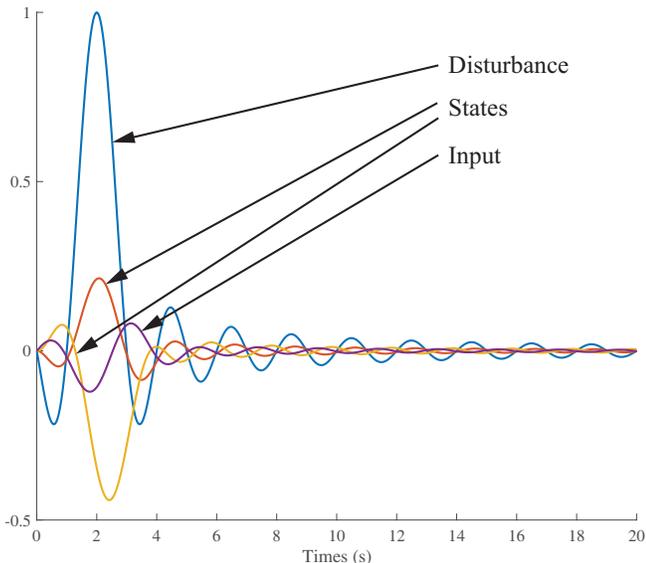}\vspace{-3mm}
\caption{Closed-loop system response to a sinc disturbance for Ex. B.3}\label{fig:ex_b3}
\end{figure}
\paragraph{Example B.3} This example is a modified version of the example in~\cite{cao_1998} ($B_2$ was modified to make the problem more difficult and regulated outputs and disturbances were added). In that work, the authors were able to find a stabilizing controller for a maximum delay of $\tau_1=.1934$ and $\tau_2=.2387$. We are able to find a controller for any $\tau_1$ and $\tau_2$. The results here are for $\tau_1=1$ and $\tau_2=2$. The closed-loop system response is illustrated in Fig.~\ref{fig:ex_b3}.
\begin{align*}
\dot x(t)=&\bmat{-1 &2\\0& 1}x(t)+\bmat{.6 &-.4\\0&0}x(t-\tau_1)+\bmat{0 &0\\0&-.5}x(t-\tau_2)+\bmat{1\\1}w(t) + \bmat{0\\1}u(t)\\
y(t)=&\bmat{1 & 0\\0 &1\\0&0}x(t)+\bmat{0\\0\\.1}u(t)
\end{align*}

\[
\hbox{\begin{tabular}{c|c|c|c|c}
$\gamma_{\min}$ & $d=1$  & $d=2$ & $d=3$  & \text{Pad\'e} \\
\hline
$\tau_1=1,\tau_2=2$ &  .6104  & .6104 & .6104 &   .6104 \\
\hline
CPU sec & 2.07 & 7.25 & 25.81  & N/A  \\
\end{tabular}}
\]

\paragraph{Example B.4} In this example, we rigorously examine the computational complexity of the proposed algorithm. We use a generalized n-D system with K delays, a single disturbance $w(t)$ and a single input $u(t)$.
\begin{align*}
\dot x(t)=&-\sum_{i=1}^{K} \frac{x(t-i/K)}{K} + \mbf{1} w(t)+ \mbf{1} u(t)\\
y(t)=&\bmat{\mbf{1}^T \\ 0}x(t)+\bmat{0\\1}
\end{align*}
where $\mbf{1}\in \R^{n}$ is the vector of all ones. The resulting computation time is listed in Table~\ref{tab:complexity}. The achieved closed-loop $H_\infty$ norms are listed in Table~\ref{tab:bounds}.
\begin{table}
  \centering
\begin{tabular}{c|c|c|c|c|c}
$K\downarrow$ $n\rightarrow$ & $1$  & $2$ & $3$ & $5$ & $10$\\
\hline
1 & .438 & .172 &  .266 &  1.24 & 17.2\\
\hline
2 & .269 & .643 &  2.932 & 17.1 & 647.2 \\
\hline
3 & .627 & 2.634 & 10.736 &  91.43 & 5170.2 \\
\hline
5 & 1.294 & 13.12 & 84.77.7 & 1877 & 65281 \\
\hline
10 & 11.41 &469.86 & 4439  & 57894  &  NA \\
\end{tabular}
  \caption{ CPU sec indexed by \# of states ($n$) and \# of delays ($K$)}\vspace{-4mm}\label{tab:complexity}
\end{table}

\begin{table}
  \centering
\begin{tabular}{c|c|c|c|c|c}
$K\downarrow$ $n\rightarrow$ & $1$  & $2$ & $3$ & $5$ & $10$\\
\hline
1 & .9235 & .9791 &  .9906 &  .9966 & .9991\\
\hline
2 & .8039 & .9379 &  .9709 & .9892 & .9973 \\
\hline
3 & .7657 & .9220 & .9630 &  .9862 & .9965 \\
\hline
5 & .7389 & .9099 & .9568 & .9838 & .9959 \\
\hline
10 & .7224 & .9020  &  .9527  & .9822  &  NA \\
\end{tabular}
  \caption{ Closed-loop norm bound indexed by \# of states ($n$) and \# of delays ($K$)}\vspace{-4mm}\label{tab:bounds}
\end{table}
As expected, these results indicate the synthesis problem is not significantly more complex that the stability test. The complexity scales as a function of $nK$ and is possible on desktop computers when $nK<50$.

\begin{figure}\vspace{-3mm}
\includegraphics[width=.5\textwidth]{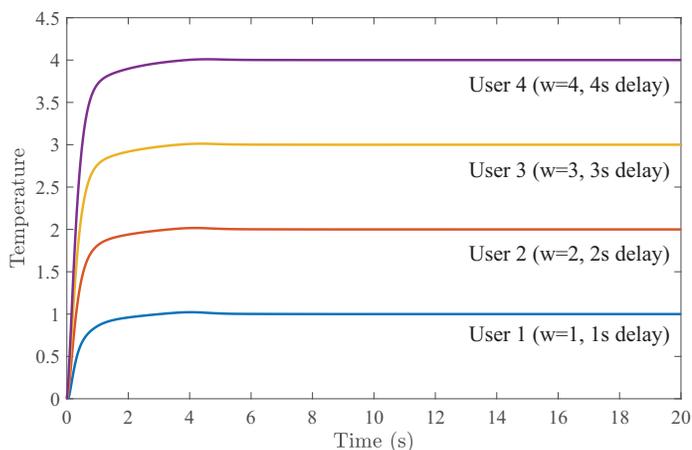}\vspace{-3mm}
\caption{A Matlab simulation of the step response of the closed-loop temperature dynamics ($T_{2i}(t)$) for System~\eqref{eqn:shower} with 4 users ($w_i$ and $\tau_i$ as indicated) coupled with the controller from Theorem~\ref{thm:synthesisLMI_MD} with closed-loop gain of $.36$}\label{fig:shower}
\end{figure}

\subsection{A Scalable Design Example with Multiple State Delays}
In this subsection, we demonstrate the scalability and potential applications of the algorithm by consider a practical problem faced in hotel management with a centralized hot-water source with multiple showering customers (a generalization of the model proposed in~\cite{peet_thesis}). Specifically, let us first consider a single user attempting to achieve a desired shower temperature by adjusting a hot-water tap. In this case, we have an significant transport delay caused by the flow of hot water from the tap to the showerhead. In modeling the dynamics, we assume that a person will adjust the tap at a rate proportional to the difference between current temperature and desired temperature and the overall flow rate is constant (i.e. does not depend on temperature). Under these assumptions, we can model the linearized water temperature dynamics at the tap as
\[
\dot T(t)=-\alpha \left(T(t-\tau)-w(t)\right)
\]
where $T$ is the water temperature and $w(t)$ is the desired water temperature. When multiple users are present and the available hot water pressure is finite, the actions of each user will affect the temperature of all other users. In a linearized model we represent this as
\[
\dot T_i(t)=-\alpha_{i} (T(t-\tau_i)-w_i(t)) -\sum_{j \neq i}\gamma_{ij}\dot T_j(t)
\]
or
\begin{align*}
\dot T_i(t)=&-\alpha_{i} \left(T(t-\tau_i)-w_i(t)\right)+ \sum_{j \neq i} \gamma_{ij} \alpha_{j}\left(T_j(t-\tau_j)-w_j(t)\right)
\end{align*}
where we have neglected products $\gamma_{ij}\gamma_{jk}$ as it is assumed these coupling coefficients are small. Even for a single user, these dynamics are often unstable if the delay is significant. For this reason, we introduce a centralized tracking control system to stabilize the temperature dynamics. Obviously, this controller can not sense the desired water temperatures, $w_i(t)$. The controller can, however, sense the tap position and the actual water temperature. We account for this by including an augmented state, $T_{1i}$ which then represents the tap position chosen by user $i$. Introducing an input into the temperature dynamics yields
\begin{align}
&\dot T_{1i}(t)=T_{2i}(t)-w_i(t) \label{eqn:shower}\\
&\dot T_{2i}(t)=-\alpha_i \left( T_{2i}(t-\tau_i) -w_i(t)\right)+ \sum_{j \neq i} \gamma_{ij} \alpha_{j}\left(T_j(t-\tau_j)-w_j(t)\right)  + u_i(t) \notag\\
 & y_i(t)  = \bmat{T_{1i}(t)\\.1u_i(t)}.\notag
\end{align}
Aggregating these dynamics into the form of Equation~\eqref{eqn:MDS}, we have
\begin{align*}
A_0&=\bmat{0 & I\\ 0 & 0},\quad A_i=\bmat{0 & 0\\ 0 & \hat A_i}\\
\hat A_i(:,i)&=\alpha_i\bmat{\gamma_{i,1}& \hdots & \gamma_{i,i-1}&-1& \gamma_{i,i-1} &\hdots & \gamma_{i,K}}^T\\
B_1&=\bmat{-I\\-\hat \Gamma +\diag(\alpha_1 \hdots \alpha_K) }\\
\hat \Gamma_{ij}&=\alpha_j\gamma_{ij}=\bmat{q_1&\hdots &q_K},\quad B_2=\bmat{0\\I}\\
C_0&=\bmat{I&0\\0&0},\;C_1=\bmat{0&0\\0&0},\; D_{1}=\bmat{0\\0},\; D_2=\bmat{0\\.1 I}
\end{align*}
%
\textbf{Optimal Control of Showering Users}\noindent

For numerical implementation with $n_u$ users, we have a system with $2n_u$ states, $n_u$ delays, $2n_u$ regulated outputs and $n_u$ control inputs. The implementation of this example is included in the accompanying code, wherein we set $\alpha_i=1$, $\gamma_{ij}=1/n$ and $\tau_i=i$. The resulting open-loop dynamics are unstable. For $n_u=4$, we obtain a closed-loop $H_\infty$ norm bound of $\gamma=.38$. For $w_i(t)=i$, the resulting closed-loop dynamics are illustrated in Figure~\ref{fig:shower} wherein convergence to the desired shower temperature is observed for all users.

\section{Conclusion}
In this paper, we have shown how the problem of optimal control of systems with multiple delays can be reformulated as a convex optimization problem with operator variables. We have proposed a parametrization of positive operators using positive matrices and verified the resulting LMIs are accurate to several decimal places when measured by the minimal achievable closed-loop $H_\infty$ norm bound. We have developed an analytic formula for the inverse of the proposed parameterized class of positive operators. Finally, we have demonstrated effective methods for real-time computation of the control inputs. Finally, we have implemented the proposed algorithms and gains and simulated the results on a realistic model with 8 states and 4 delays.
\section*{Acknowledgment}
This work was supported by the National Science Foundation under grants No. 1301660, 1538374 and 1739990.

\bibliographystyle{IEEEtran}
\bibliography{peet_bib,delay,NSF_CAREER_bib2011,LMIs}

\begin{IEEEbiography}{Matthew M. Peet}
received the B.S. degree in physics and in aerospace engineering from the University of Texas, Austin, TX, USA, in 1999 and the M.S. and Ph.D. degrees in aeronautics and astronautics from Stanford University, Stanford, CA, in 2001 and 2006, respectively. He was a Postdoctoral Fellow at INRIA, Paris, France from 2006 to 2008. He was an Assistant Professor of Aerospace Engineering at the Illinois Institute of Technology, Chicago, IL, USA, from 2008 to 2012. Currently, he is an Associate Professor of Aerospace Engineering at Arizona State University, Tempe, AZ, USA. Dr. Peet received a National Science Foundation CAREER award in 2011.
\end{IEEEbiography}

\end{document}